\documentclass[fleqn,usenatbib]{mnras}

\usepackage[english]{babel} 
\usepackage[utf8]{inputenc} 
\usepackage[T1]{fontenc}    
\usepackage{ae,aecompl}
\usepackage{pgf,pgfarrows,pgfnodes,pgfautomata,pgfheaps}
\usepackage{graphicx}       
\usepackage{natbib}         
\usepackage{url}            
\usepackage{grffile}        
\usepackage{mathtools}      
\usepackage{multirow}       
\usepackage{xspace}         
\usepackage{booktabs}       

\usepackage{txfonts}
\usepackage{amsmath,amssymb,amsxtra,amsfonts}   

\usepackage{color}         
\definecolor{gold}{rgb}{1,0.80,0}
\definecolor{orange}{rgb}{1,0.5,0}
\definecolor{midgray}{gray}{0.3}
\definecolor{lblue}{rgb}{0,0.2,0.6}
\definecolor{dgreen}{rgb}{0.1,0.6,0.3}
\definecolor{purple}{rgb}{0.5019607843137255,0.0,0.5019607843137255}

\renewcommand{\arcsec}{\mbox{$^{\prime\prime}$}\xspace}  

\newcommand{\be}{\begin{equation}}
\newcommand{\ee}{\end{equation}}

\newcommand{\ba}{\begin{align}}
\newcommand{\ea}{\end{align}}

\newcommand{\defeq}{\vcentcolon=}

\newcommand{\Msun}{\ensuremath{M_\odot}\xspace}

\newcommand{\Mstar}{\ensuremath{M_\ast}\xspace}

\newcommand{\oh}{\ensuremath{12+\log({\rm O/H})}\xspace}

\newcommand{\Hunit}{\ensuremath{\rm km~s^{-1}~Mpc^{-1}}\xspace}

\newcommand{\Ha}{\textrm{H}\ensuremath{\alpha}\xspace}
\newcommand{\Hb}{\textrm{H}\ensuremath{\beta}\xspace}
\newcommand{\Hg}{\textrm{H}\ensuremath{\gamma}\xspace}
\newcommand{\Hd}{\textrm{H}\ensuremath{\delta}\xspace}
\newcommand{\HII}{\textrm{H}\textsc{ii}\xspace}

\newcommand{\OI}{[\textrm{O}~\textsc{i}]\xspace}
\newcommand{\OII}{[\textrm{O}~\textsc{ii}]\xspace}
\newcommand{\OIII}{[\textrm{O}~\textsc{iii}]\xspace}

\newcommand{\NII}{[\textrm{N}~\textsc{ii}]\xspace}
\newcommand{\SII}{[\textrm{S}~\textsc{ii}]\xspace}
\newcommand{\NeIII}{[\textrm{Ne}~\textsc{iii}]\xspace}

\newcommand{\emc}{\textsc{Emcee}\xspace}

\newcommand{\grzl}{\textsc{Grizli}\xspace}

\newcommand{\bagp}{\textsc{Bagpipes}\xspace}

\newcommand{\jwst}{\textit{JWST}\xspace}

\newcommand{\glass}{\textit{GLASS}\xspace}

\def\ie{i.e.\xspace}
\def\eg{e.g.\xspace}

\newcommand{\Om} {\ensuremath{\Omega_{\rm{m}}}\xspace}

\newcommand{\Ol} {\ensuremath{\Omega_{\Lambda}}\xspace}

\defcitealias{Bian:2018km}{B18}
\defcitealias{Nakajima:2022ApJS..262....3N}{N22}
\defcitealias{Cataldi:2025A&A}{Ca25}
\defcitealias{Sanders:2025arXiv}{S25}
\defcitealias{Chakraborty:2025ApJ}{Cha25}

\newcommand{\Bian}{\citetalias{Bian:2018km} }
\newcommand{\Nakajima}{\citetalias{Nakajima:2022ApJS..262....3N} }
\newcommand{\Cataldi}{\citetalias{Cataldi:2025A&A} }
\newcommand{\Sanders}{\citetalias{Sanders:2025arXiv} }
\newcommand{\Chakra}{\citetalias{Chakraborty:2025ApJ} }

\usepackage{newtxtext,newtxmath}
\usepackage{orcidlink}     

\usepackage[T1]{fontenc}
\usepackage{pdflscape}
\usepackage{lineno}
\DeclareRobustCommand{\VAN}[3]{#2}
\let\VANthebibliography\thebibliography
\def\thebibliography{\DeclareRobustCommand{\VAN}[3]{##3}\VANthebibliography}

\usepackage{graphicx}	
\usepackage{amsmath}	

\long\def\rp#1{#1}

\title[NIRISS MZR and FMR at $z=1.1-3.4$]{
Revisiting the mass metallicity relation and the fundamental metallicity relation of dwarf galaxies at cosmic noon with NIRISS
}

\author[X. He et al.]{%
Xianlong He \orcidlink{0000-0002-1336-5100}$^{1,2,3}$, 
Zihao Li \orcidlink{0000-0001-5951-459X}$^{4,5,6}$ ,
Xin Wang \orcidlink{0000-0002-9373-3865}$^{1,7,8}$ \thanks{Corresponding Author E-mail: xwang@ucas.ac.cn},
Zheng Cai \orcidlink{0000-0001-8467-6478}$^{6}$,
Tucker Jones \orcidlink{0000-0001-5860-3419}$^{9}$,
\newauthor
Tommaso Treu \orcidlink{0000-0002-8460-0390}$^{3}$ ,
Benedetta Vulcani \orcidlink{0000-0003-0980-1499}$^{10}$,
Matthew A. Malkan \orcidlink{0000-0001-6919-1237}$^{3}$,
and Karl Glazebrook \orcidlink{0000-0002-3254-9044}$^{11}$
\\
$^{1}$School of Astronomy and Space Science, University of Chinese Academy of Sciences (UCAS), Beijing 100049, China\\
$^{2}$Department of Astronomy, School of Physics and Technology, Wuhan University, Wuhan 430072, People's Republic of China\\
$^{3}$Department of Physics and Astronomy, University of California, Los Angeles, 430 Portola Plaza, Los Angeles, CA 90095, USA\\
$^{4}$Cosmic Dawn Center (DAWN), Denmark\\
$^{5}$Niels Bohr Institute, University of Copenhagen, Jagtvej 128, DK-2200 Copenhagen N, Denmark\\
$^{6}$Department of Astronomy, Tsinghua University, Beijing 100084, People's Republic of China\\
$^{7}$National Astronomical Observatories, Chinese Academy of Sciences, Beijing 100101, China  \\
$^{8}$Institute for Frontiers in Astronomy and Astrophysics, Beijing Normal University, Beijing 102206, China  \\
$^{9}$ Department of Physics and Astronomy, University of California Davis, 1 Shields Avenue, Davis, CA 95616, USA \\
$^{10}$ INAF Osservatorio Astronomico di Padova, vicolo dell'Osservatorio 5, 35122 Padova, Italy\\
$^{11}$ Centre for Astrophysics and Supercomputing, Swinburne University of Technology, PO Box 218, Hawthorn, VIC 3122, Australia \\
}

\date{Accepted XXX. Received YYY; in original form ZZZ}

\pubyear{\the\year{}}

\begin{document}
\label{firstpage}
\pagerange{\pageref{firstpage}--\pageref{lastpage}}
\maketitle

\begin{abstract}
We extend the stellar-mass gas-phase metallicity relation (MZR) at $z = 1.1-3.4$ down to the extremely low-mass regime using 183 galaxies with $\log(M_*/M_\odot) = 6.3-10.2$, based on deep JWST/NIRISS slitless spectroscopy from the NGDEEP program. 
The derived MZR is in excellent agreement with our previous result from 50 galaxies in the GLASS-JWST sample, underscoring the robustness and universality of this relation. 
Together, these datasets constitute the largest sample of dwarf galaxies yet obtained with NIRISS. 
The observed MZR slope, $\beta\simeq0.24\pm0.03$, remains constant across nearly four orders of magnitude in stellar mass. 
Analytical modeling of the metal-loading factor of outflows ($\zeta_\textrm{out}$) indicates that, at $M_*\lesssim10^8M_\odot$, $\zeta_\textrm{out}$ becomes progressively less dominant than the gas fraction ($\mu_\textrm{gas}$) in regulating the MZR slope. 
Using this enlarged NIRISS sample, we further test the existence of the fundamental metallicity relation (FMR). 
We find no robust evidence for an additional SFR dependence beyond the MZR, nor any reduction in metallicity scatter when SFR is included. 
Examination of systematic uncertainties in \oh, $M_*$ and SFR suggests that the MZR slope ($\beta\sim0.22$) is robust, and that different assumptions about the strong-line calibrations or star-formation history (SFH) of the galaxies change the slope by less than 1-$\sigma$. 
At the current depth of the NIRISS data, evidence for an FMR among high-redshift dwarf galaxies remains inconclusive, highlighting the need for larger samples, and deeper observations.

\end{abstract}




\section{\rp{Introduction}} 
\label{sec:intro}

Gas-phase metallicity is a key observational tracer of the baryon cycle in galaxies.
Heavy elements are produced by successive generations of stars and returned to the interstellar medium through stellar winds and supernova explosions.
The amount of metals retained in the gas phase is then regulated by the balance between metal production, gas accretion, star formation, feedback-driven outflows, and gas recycling \citep{Finlator:2008MNRAS.385.2181F,Dave:2012MNRAS.421...98D,Lilly:2013ko,Dekel:2014jm,Peng:2014hn}.
Therefore, elemental abundances provide a fossil record of the past star-formation and gas-flow histories of galaxies, and offer a powerful way to connect galaxy growth with feedback and chemical enrichment across cosmic time \citep{Maiolino:2019vq}.
In practice, gas-phase oxygen abundance, commonly expressed as $12+\log(\mathrm{O/H})$, is usually inferred from rest-frame optical emission lines, making metallicity measurements directly tied to the nebular conditions of star-forming regions \citep{Kewley:2019kf}.

On galaxy-wide scales, the regulation of metal enrichment by the baryon cycle is encoded in empirical relations among stellar mass ($M_*$), gas-phase metallicity (\oh), and star-formation rate (SFR).
\rp{
The best-established of these is the stellar mass--metallicity relation (MZR) \citep[e.g.,][]{Curti:2025arXiv}, in which more massive galaxies are generally more metal-rich.
}
In the local Universe, this relation was established with large spectroscopic samples of star-forming galaxies \citep{Tremonti:2004ed}, and has since been revisited using direct-$T_{\rm e}$-anchored abundance scales \citep{Andrews:2013dn,Curti:2020MNRAS.491..944C}.
The local MZR is relatively flat at high stellar masses, while below the turnover mass of $M_*\lesssim10^{10.5}M_\odot$ it is commonly described as a power law, although the inferred slope (e.g., $\beta\sim0.2-0.4$) and normalization depend on the adopted calibration \citep{Kewley:2008be}.
\rp{
Direct-$T_{\rm e}$ measurements of nearby dwarf galaxies further show that the MZR extends into the low-mass regime \citep[e.g.,][]{Berg:2012ApJ}.
}
A large number of theoretical models have been proposed to explain the origin of the MZR, including analytical gas-regulator models, semi-analytical models, and hydrodynamical simulations \citep{Finlator:2008MNRAS.385.2181F,Erb:2008di,Peeples:2011ew,Dave:2012MNRAS.421...98D,Lilly:2013ko,Lu:2014kl,Ma:2016gw,Dave:2017gd,Torrey:2019MNRAS.484.5587T,Bassini:2024MNRAS}.
In these models, the MZR reflects the balance between metal production, gas accretion, star formation, feedback-driven outflows, and gas recycling.
Its slope and normalization are therefore sensitive to gas fraction, gas dilution, and the mass and metal loading of galactic outflows \citep{Peeples:2011ew,Dave:2013MNRAS,Guo:2016ApJ,Sanders:2021ga,Tortora:2022A&A}.
Although metallicity measurements introduce systematic uncertainties, the relative behaviour of the MZR across stellar mass can still provide insight into the physical mechanisms regulating metal retention and metal loss.

The residual scatter around the MZR is not purely random, but shows a secondary dependence on SFR.
At fixed stellar mass, galaxies with higher SFRs tend to have lower metallicities, motivating the fundamental metallicity relation (FMR), a three-dimensional relation among $M_*$, gas-phase metallicity, and SFR \citep{Mannucci:2010MNRAS.408.2115M,Lara-Lopez:2010A&A,Andrews:2013dn}.
The local FMR was originally reported to have a small residual scatter of $\sim0.05$ dex in metallicity \citep{Mannucci:2010MNRAS.408.2115M}, \rp{
and has subsequently been re-examined on a fully $T_{\rm e}$-based abundance scale \citep{Curti:2020MNRAS.491..944C}.
}
Physically, the SFR dependence is often interpreted in terms of variations in gas fraction, gas accretion, metal dilution, and metal-loaded outflows at fixed stellar mass \citep{Finlator:2008MNRAS.385.2181F,Dave:2012MNRAS.421...98D,Lilly:2013ko}.
Thus, the MZR and FMR provide complementary projections of the same underlying connection between galaxy growth, star formation, and chemical enrichment.

At higher redshifts, the MZR and FMR have been tested with rest-frame optical spectroscopy during the epoch of peak cosmic star formation.
Before JWST, near-infrared ground-based surveys and grism/slitless spectroscopy extended MZR measurements to $z\sim1-3$ and showed that galaxies at fixed stellar mass are typically more metal-poor than their local counterparts \citep{Erb:2006kn,Maiolino:2008A&A...488..463M,Zahid:2014ia,Sanders:2015gk,Sanders:2021ga,Henry:2021ju}.
These studies also provided the first tests of whether the local FMR persists at cosmic noon, although the results remain sensitive to sample selection, SFR indicators, and metallicity calibrations \citep{Wuyts:2012gb,Zahid:2014ia,Sanders:2018ApJ,Cresci:2019A&A,Sanders:2021ga,KorhonenCuestas:2025ApJ}.
The advent of JWST has extended these studies to lower stellar masses and higher redshifts, enabling measurements of the MZR and the $M_*-Z-$SFR relation out to $z\gtrsim4-10$ \citep[e.g.][]{Curti:2023MNRAS,Nakajima:2023arXiv230112825N,Chemerynska:2024arXiv,Sarkar2024arXiv240807974S,Stanton:2026MNRAS}.
Rather than simply increasing the number of high-redshift measurements, these data have highlighted the importance of homogeneous line-flux measurements, sample completeness, and metallicity calibrations when comparing MZR/FMR results across cosmic time.
Recent JWST auroral-line studies have begun to establish direct-$T_{\rm e}$-anchored calibrations at high redshift and at cosmic noon \citep{Sanders:2024ApJ,Cataldi:2025A&A,Chakraborty:2025ApJ,Gimenez-Alcazar:2026arXiv}, but current samples are still limited in size and often occupy restricted ranges in stellar mass, excitation, and metallicity.
Consequently, whether the apparent evolution of the MZR/FMR reflects genuine redshift evolution, calibration systematics, or the different stellar-mass regimes probed at high redshift remains an open question.

However, a key implicit limitation in many high-redshift MZR/FMR studies is that they do not probe the same stellar-mass regime as the local samples on which the MZR and FMR are usually calibrated \citep{Laseter:2025arXiv}, since the shallow gravitational potential wells of dwarf galaxies make them especially sensitive to stellar feedback, gas accretion, and metal-loaded outflows \citep{Hodge:1971ARA&A,Mateo:1998ARA&A,Tolstoy:2009ARA&A,Simon:2019ARA&A}.
\rp{
In the local Universe, direct-$T_{\rm e}$ measurements have established a benchmark MZR for dwarf galaxies; for example, \citet{Berg:2012ApJ} measured \OIII~$\lambda4363$ in 31 low-luminosity galaxies and derived a low-mass MZR slope of $0.29\pm0.03$ with a scatter of $0.15$ dex.
}
Beyond the local Universe, \citet{Henry:2013gx} provided one of the first robust low-mass MZR measurements at $1.3\lesssim z\lesssim2.3$, using stacked HST/WFC3 grism spectra of 83 emission-line galaxies down to $M_*\sim10^8M_\odot$.
More recently, slitless and integral-field studies have extended such measurements to larger or lower-mass samples, showing that low-mass galaxies at $z\sim1-2$ preserve a positive MZR while lying below the local relation \citep{Guo:2016ApJ,Henry:2021ju,Pharo:2023ApJ,Revalski:2024arXiv}.
Nevertheless, the evolution of the low-mass MZR slope and normalization, and its possible secondary dependence on SFR, remain poorly constrained at cosmic noon, especially below $M_*\lesssim10^9M_\odot$.
This leaves open whether the apparent behaviour of the FMR at high redshift reflects genuine redshift evolution, or instead the fact that high-redshift samples increasingly probe a low-mass regime where the local FMR itself is not well anchored.


Taken together, these recent studies leave a key regime insufficiently explored: dwarf galaxies at cosmic noon.
On the one hand, systematic FMR studies at $z\sim2$ have mainly focused on more massive galaxies; for example, \citet{KorhonenCuestas:2025ApJ} found no robust evidence for an extension of the local FMR to a KBSS sample at $z\sim2.3$ with $\log(M_*/M_\odot)\simeq9-11$.
On the other hand, recent work on the low-mass regime has suggested that apparent high-redshift FMR offsets may partly reflect the poor anchoring of the local FMR below $\log(M_*/M_\odot)\lesssim8.5-9$, rather than redshift evolution alone \citep{Laseter:2025arXiv}.

In this work, we bridge these two directions by measuring the MZR and testing the FMR for a large sample of low-mass galaxies at $z\sim1-3.5$ using JWST/NIRISS spectroscopy.
Building on our previous GLASS-JWST analysis \citep{He:2024ApJL}, we extend the sample with NGDEEP observations and quantify how metallicity calibrations, stellar-mass estimates, and SFR indicators affect the inferred MZR and FMR.
This analysis provides a benchmark for interpreting current and future JWST measurements of chemical scaling relations in low-mass galaxies at cosmic noon and beyond.
We present a measurement of the MZR and FMR using the NIRISS data from a sample of  183 field galaxies for $\log(M_*/M_\odot)=7-9$ at $z=1-3$ as presented in Fig.~\ref{fig:mzr}, \& ~\ref{fig:fmr}. 
In Sect.~\ref{sec:data}, we describe the data acquisition and galaxy sample analyzed in this work.
In Sect.~\ref{sec:measure}, we demonstrate our method for extracting metallicity and stellar mass. 
We discuss the results in Sect.~\ref{sec:result} and summarize the main conclusions in Sect.~\ref{sec:conclude}.
The AB magnitude system, the standard concordance cosmology ($\Om=0.3, \Ol=0.7$, $H_0=70\,\Hunit$), and the \citet{Chabrier:2003ki} Initial Mass Function (IMF) are adopted.  
The metallic lines are denoted in the following manner, if presented without wavelength: $\OII~\lambda\lambda3727,3730\defeq\OII,\NeIII~\lambda3869\defeq\NeIII,\Hg~\lambda4342\defeq\Hg, \Hb~\lambda4863\defeq\Hb, \OIII~\lambda5008\defeq\OIII, \Ha~\lambda6564\defeq\Ha, \SII~\lambda\lambda6716,6731\defeq\SII$.

\section{Observations} 
\label{sec:data}

We use JWST/NIRISS Wide Field Slitless Spectroscopy (WFSS) observations from the Next Generation Deep Extragalactic Exploratory Public survey 
\citep[NGDEEP; Proposal ID \#2079;][]{Bagley:2023arXiv}.
This treasury survey targets the Hubble Ultra Deep Field (HUDF) using NIRISS over a field of view of $133\arcsec\times133\arcsec$ using the GR150R and GR150C grisms, with the goal of measuring metallicities and star-formation rates of low-mass galaxies across the peak epoch of cosmic star formation.
These public data allow us to enlarge our previous \glass-\jwst sample \citep{He:2024ApJL} by a factor of approximately 3 and to extend the analysis toward lower stellar masses.
In this work, we analyze the first-epoch observations obtained between 2023 January 31 and February 2.
The total exposure times in the grisms are 95 ks in F115W, 43 ks in F150W, and 32 ks in F200W, together with direct imaging in the same filters with total integration times of 5.4, 1.7, and 1.7 ks, respectively \citep{Shen:2023arXiv}.
The observations provide low-resolution spectra with $R\equiv\lambda/\Delta\lambda\sim150$ for all objects in the field of view, with continuous wavelength coverage over $\lambda\simeq1.0-2.2\,\mu$m.
This wavelength range includes some or all of the strong rest-frame optical emission lines \OII, \NeIII, \Hg, \Hb, \OIII, \Ha, and \SII\ over $z=1.10$--$3.43$, as summarized in Tab.~\ref{tab:stack}.
We reduce the NIRISS WFSS data using the Grism Redshift \& Line software \citep[\grzl;][]{brammer_gabriel_2023_7712834}.



For the galaxy sample in the NGDEEP field, we used the public photometry catalogue from the DAWN JWST Archive (DJA).
Basic details of the NIRCam data reduction are presented by \citet{Valentino:2023ApJ}.
The list of programs included in the archive so far is summarized in this \grzl document\footnote{\url{https://dawn-cph.github.io/dja/imaging/v7/}}.
All reduced images in 8 HST/WFC3-IR (F105W, F110W, F125W, F140W, F160W, F606W, F814W, F850LP), 6 HST/ACS-WFC (F435W, F475W, F606W, F775W, F814W, F850LP), 14 JWST NIRCam (F090W, F115W, F150W, F182M, F200W, F210M, F277W, F335M, F356W, F410M, F430M, F444W, F460M, F480M), and 3 JWST NIRISS (F115W, F150W, F200W) filters are used if available.
This photometric catalogue of 31 bands with an observed-frame wavelength coverage of $\lambda \in [0.4, 5.0] \mu$m at $z\in[1.10, 3.43]$, allows very robust stellar mass estimations.

\begin{figure*}
    \centering
    \includegraphics[width=\textwidth]{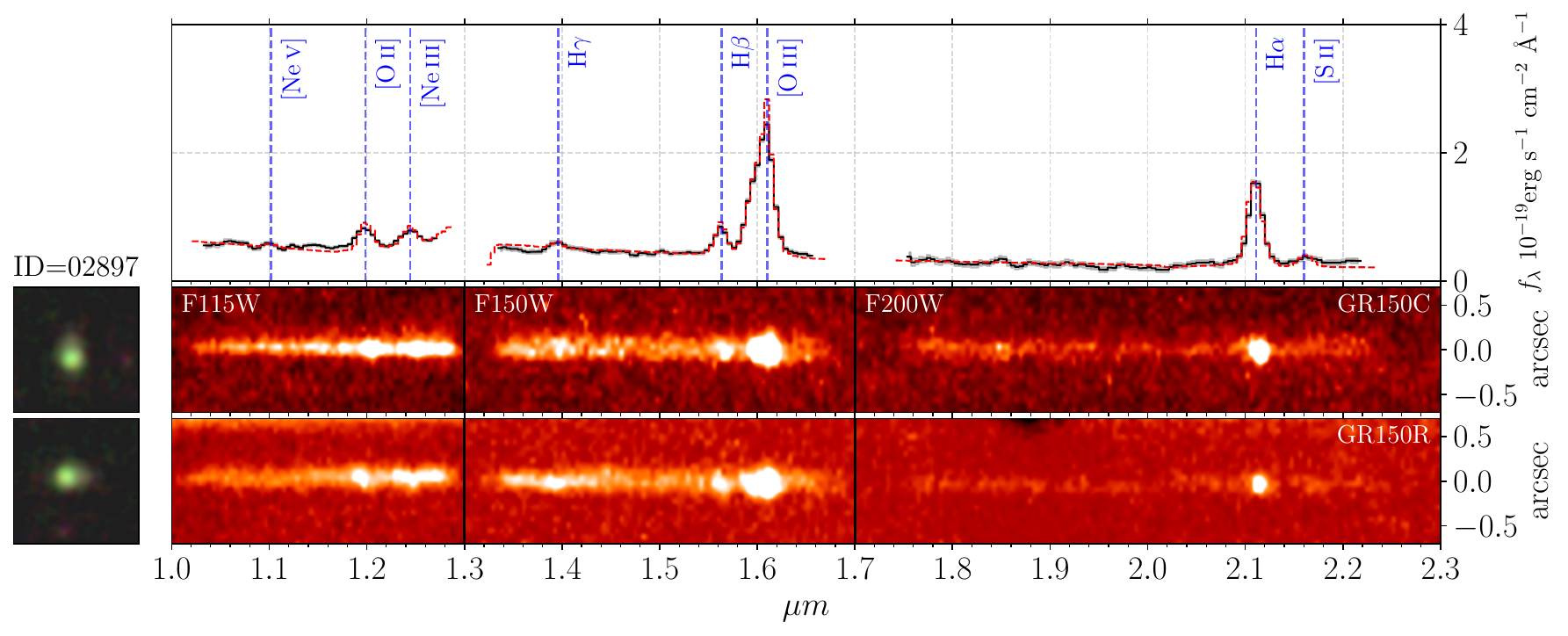}
    \caption{Example data for one source (ID=02897) at redshift $z=2.2155$ with $\log(M_*/M_\odot)=8.132$, including photometric (left) and spectroscopic (right) observations. 
    The left two panels show the pseudo-color image consisting of 3 bands F115W, F150W, F200W from NIRISS pre-image, aligned with the corresponding dispersion direction of spectroscopy.
    The observational frame 2D NIRISS spectra in 3 bands of 2 dispersion GR150C, GR150R are shown in the bottom right panels. The upper right panel shows the 1D extracted spectrum, marking the emission line position in the observational frame. The data and error are displayed as black lines and gray belts, while the fitted models are represented by red dashed lines.
    }
    \label{fig:indivi_example}
\end{figure*}

\section{Analysis}\label{sec:measure}

In this section, we present the measurements of the physical properties derived from spectroscopy and photometry for individual galaxies and stacked spectra, following our previous work \citep{Jones:2015AJ....149..107J,Wang:2016um,Wang:2019cf,Wang:2020bp,Wang:2022ApJ...926...70W}.
Interested readers are referred to \citet{He:2024ApJL} for further details.

\subsection{Grism Redshift and Emission-line Flux}
\label{subsec:gri}

We utilize the Grism Redshift and Line Analysis software \textsc{Grizli} \citep{brammer_gabriel_2023_7712834} to reduce NIRISS data using the standard JWST pipeline version 1.9.4 and the \texttt{jwst\_1090.pmap} reference file. The detailed procedures are largely described by \cite{RobertsBorsani:2022ApJ...938L..13R}. 
Briefly, \grzl analyzes the paired direct imaging and grism exposures through forward modeling, and yields contamination subtracted 1D \& 2D grism spectra, the best-fit spectroscopic redshifts and the emission line fluxes. 
The forward modeling procedure begins with the construction of a one-dimensional (1D) spectrum, consisting of several Gaussian-shaped nebular emission lines at a given redshift and a continuum superimposed by four sets of empirical spectra \citep{Brammer:2008gn,Erb:2010iy,Muzzin:2013is,Conroy:2012bl}.
The modeled 1D spectrum is utilized to generate a 2D model spectrum based on the grism sensitivity and dispersion function, similar to the 'fluxcube' model produced by the aXe software \citep{2009PASP..121...59K}.
This 2D forward-modeled spectrum is then compared to the data and a global $\chi^2$ calculation is performed to determine the best-fit superposition coefficients for both the continuum templates and Gaussian amplitudes corresponding to the best-fit emission line fluxes.
We refer interested readers to Appendix A of \citet{Wang:2019cf}, for a full description of the redshift fitting procedure.

We obtain a parent sample of 1181 sources  with NIRISS spectra coverage, and the spectra can be fitted by \grzl.
Since the fitting procedure in the first round uses a uniform prior on redshift $z\in[0,9]$, the $\chi^2$ are likely reaching the local minimum, giving spurious redshift solutions. 
To build a complete emitter catalog, we additionally perform a visual inspection on both the 1D and 2D spectra, both removing false \OIII emitter candidates and recovering missing ones. 
The emission lines of a good candidate can be identified in both dispersion directions, as an example shown in the bottom right panel in Fig.~\ref{fig:indivi_example}. We mark the locations of the likely \OIII emission lines and estimate the redshifts for each source. We then apply the redshift fitting procedure again, given the visual-inspected redshift as the redshift prior. After we have fitted all the \OIII candidates in the second round, 
a total of 223 sources are identified reliably in the redshift range of $z\in[1.03, 1.55]\cup[1.70, 2.35]\cup [2.54, 3.44]$.
To minimally ensure the reliable estimation of metallicity and SFR, we also need \Hb detection.
There are 186 galaxies with $\mathrm{SNR}>2$ of \Hb and no other emission line criteria (e.g., SNR of \OIII) are used for selection, to avoid potential metallicity bias.
After the 3 possible AGN exclusions in Sect.~\ref{subsec:mass}, the remaining 183 sources make up our final sample with 38, 70, 75 galaxies at each range $z\in[1.10, 1.54]\cup[1.76, 2.32]\cup [2.61, 3.43]$, respectively, showing prominent nebular emission lines. 
We exhibit an example of a galaxy in Fig.~\ref{fig:indivi_example} with its photometric and spectroscopic data from NIRISS.
In addition to spectra and redshifts, the line fluxes $(f_i^\text{o}, \sigma_i^\text{o})$ (covering \OII, \NeIII, \Hd, \Hg, \Hb, \OIII, \Ha, and \SII if available, the same as in \citealt{Henry:2021ju}) 2D forward modeled by \textsc{Grizli}, are important output for metallicity estimation in Sect.~\ref{subsec:metal}. 
In this paper, we focus only on the integrated line flux and its corresponding global metallicity of each galaxy. Spatially resolved metallicity has been explored in our companion work \citep{Li:2025arXivb}.

\subsection{Gas-phase metallicity and Star Formation Rate} 
\label{subsec:metal}

\rp{We estimate gas-phase metallicities using strong-line diagnostics, which have been widely applied to large samples of star-forming galaxies \citep[e.g.][]{Tremonti:2004ed,Raptis:2025arXiv}.
We can only rely on bright rest-frame optical emission lines, since the faint auroral lines, such as [O~{\sc iii}]~$\lambda4363$ required in the direct Te method, are generally not detectable with sufficient significance in our NIRISS spectra.
}
As our fiducial metallicity calibration, we adopt the $\mathrm{O}_3$--$\mathrm{O}_2$  (or defined as $\mathrm{R}_3$--$\mathrm{R}_2$ in some works) diagnostic set prescribed by 
\citet[][hereafter \Bian]{Bian:2018km}, as detailed in \citet[][Tab.~1]{He:2024ApJL}. 
\rp{
This calibration is based on local analogs of $z\sim2$ star-forming galaxies and has become one of the commonly adopted empirical calibrations for cosmic-noon metallicity studies \citep[e.g.,][]{Jain:2026ApJ,Yang:2026ApJ,Li:2025ApJS} facilitating direct comparison as shown in Fig.~\ref{fig:mzr}.
}

Following our previous series of works \citep{Jones:2015AJ....149..107J,Wang:2016um,Wang:2019cf,Wang:2020bp,Wang:2022ApJ...926...70W,He:2024ApJL}, we apply a Bayesian inference framework to the observed line fluxes and uncertainties, $(f_i^{\rm o},\sigma_i^{\rm o})$, measured in Sect.~\ref{subsec:gri}.
\rp{This framework jointly constrains the gas-phase metallicity, nebular dust attenuation, and intrinsic (de-reddened) H$\beta$ flux, denoted by $(\oh,A_V,f_{\Hb})$.
Unlike the conventional approach of converting a single line ratio directly into metallicity \citep[e.g.][]{Sanders:2021ga}, our method simultaneously uses multiple emission lines $(f_i, \sigma_i)$ and evaluates the likelihood in terms of the individual line fluxes rather than precomputed flux ratios $(R_{ij}:=f_i/f_j, \sigma_{R,ij})$.
This avoids propagating the uncertainty of a low-SNR denominator line, such as H$\beta$, into all related ratios, which would otherwise dilute the constraining power of high-SNR bright.
The method therefore naturally gives greater weight to the most informative high-SNR lines (\eg, \OII, \OIII) while marginalizing over faint (\eg, \Hb) or non-detected lines (\eg, \Hg), as well as over the nuisance parameters $A_V$ and $f_{\Hb}$.
This allows us to obtain more stable metallicity constraints even when some individual lines have low significance, as tested in Sect.~3.6 of \citet{Wang:2020bp} for \Hb with $\text{SNR}\sim1.1$.
}

The Markov Chain Monte Carlo (MCMC) sampler \emc software \citep{ForemanMackey:2013io} is employed to sample the likelihood $\mathcal{L}\propto\exp{(-\chi^2/2)}$ with:
\begin{equation}
    \chi^2 := \sum_{i}^\mathrm{EL} \frac{\left(f_i-R_{i,\mathrm{cal}}\cdot f_{\Hb}\right)^2}{\left(\sigma_{i}\right)^2 + \left(\sigma_{i,\mathrm{cal}}\cdot f_{\Hb}\right)^2} . 
    \label{eq:chi_square}
\end{equation}
The intrinsic flux and uncertainty $(f_i, \sigma_i)$ for each line ratio diagnostic $i$ , are inferred from observed values $(f_i^\text{o}, \sigma_i^\text{o})$ correcting for dust attenuation by parameter $A_V$ using the \citet{Cardelli:1989ApJ} extinction law. 
The line flux ratio $R_{i,\mathrm{cal}}$ is empirically calibrated by a polynomial as a function of metallicity: $\log{R} = \sum_{j=0}^n c_{j} \cdot (x)^j, x:=\oh$, along with its intrinsic scatter $\sigma_{i,\mathrm{cal}}. $ 
For flux ratio calibrations that do not use \Hb as the denominator (e.g.,  \NeIII/\OIII), the free parameter $f_{\Hb}$ in Eq.~\ref{eq:chi_square} needs to be replaced by the de-reddened corresponding line (e.g., $f_{\OIII}$), and one more term of uncertainty (e.g., $\left(\sigma_{\text{O3}}/f_{\text{O3}}\right)^2\cdot R_\text{Ne3O3}^2$) needs to be added to the denominator of $\chi^2$, due to the law of propagation of uncertainties.

\begin{figure*}
    \centering
    \includegraphics[width=\textwidth]{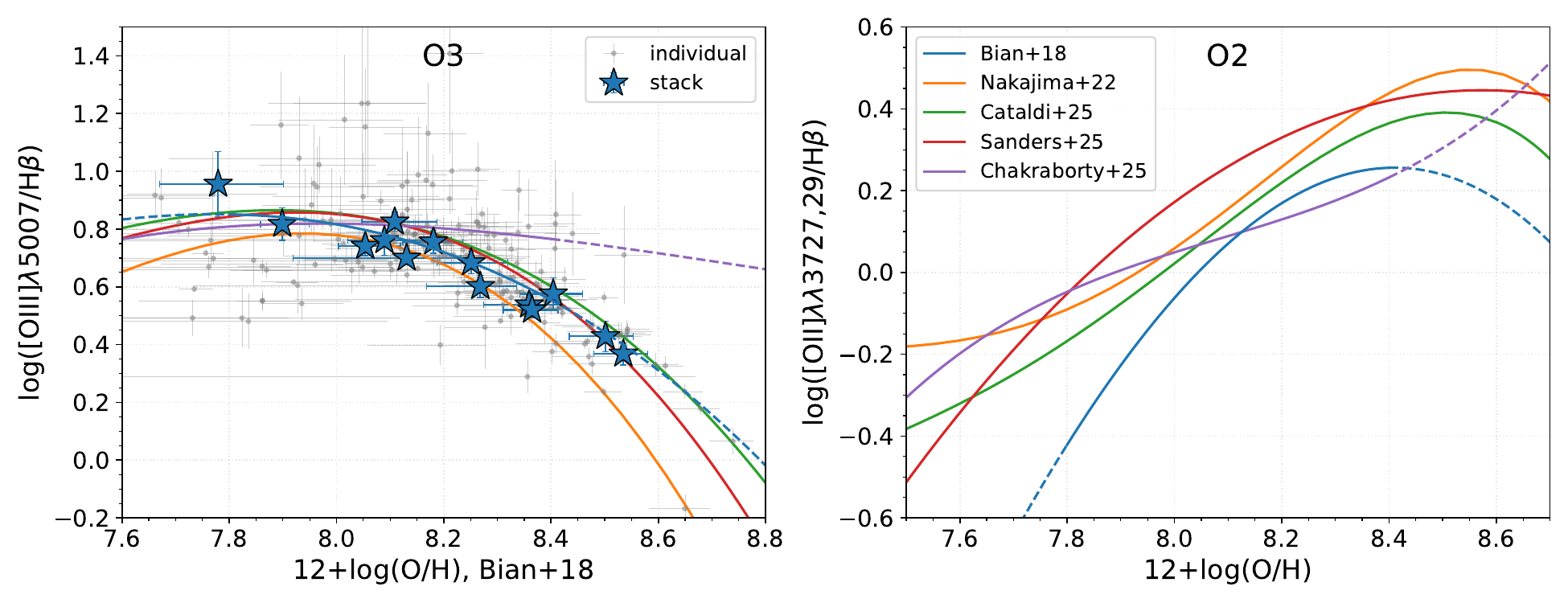} 
    \caption{\rp{
    Representative O3:=\OIII$\lambda$5007/\Hb (left) and O2:=\OII$\lambda\lambda$3727,29/\Hb (right) calibrations adopted in this work.
    The selected calibrations span local-analog \citep{Bian:2018km}, low-metallicity/high-excitation \citep{Nakajima:2022ApJS..262....3N}, cosmic-noon \citep{Cataldi:2025A&A}, and broad high-redshift empirical frameworks \citep{Sanders:2025arXiv,Chakraborty:2025ApJ}.
    Solid curves indicate the nominal metallicity ranges over which the calibrations are defined, while dashed curves show the modest extrapolations beyond these ranges adopted for comparison.
    In the left panel, we also illustrate how our Bayesian metallicity estimates compare with the fiducial \Bian single-diagnostic \OIII/\Hb calibration.
    Since our Bayesian metallicities are inferred from multiple emission-line diagnostics while marginalizing over dust attenuation and intrinsic H$\beta$ flux, modest deviations from the single-ratio curve are expected.
    }}
    \label{fig:calib}
\end{figure*}

\rp{
Many oxygen-based strong-line metallicity calibrations have been developed for star-forming galaxies \citep[e.g.][]{Maiolino:2008A&A...488..463M,Jones:2015ApJ...813..126J,Bian:2018km,Nakajima:2022ApJS..262....3N},
and the increasing number of JWST-era high-redshift auroral-line detections has enabled new empirical calibrations based directly on high-redshift galaxies \citep[e.g.][]{Cataldi:2025A&A,Sanders:2025arXiv,Chakraborty:2025ApJ}.
These recent developments have been systematically compiled and compared by \citet{Rosales-Ortega:2026arXiv} with their own calibrations.
Since the adopted calibration directly affects both the inferred metallicities and the resulting MZR/FMR measurements, we select several representative calibrations, shown in Fig.~\ref{fig:calib}, and quantify their impact in Sect.~\ref{subsec:calib}.
}
In Fig.~\ref{fig:calib}, we use \OIII/\Hb\ as a representative single-ratio projection to illustrate how our Bayesian metallicity estimates compare with the single-diagnostic calibration, since \OIII and \Hb are close in wavelength and are therefore only weakly affected by differential dust attenuation.
In the conventional single-diagnostic approach \citep[e.g.,][]{Li:2023ApJL}, a measured line ratio is converted directly into metallicity according to the calibration curve after dust correction by $A_V$.
By contrast, our Bayesian framework simultaneously combines multiple line diagnostics and jointly infers \oh, $A_V$, and the intrinsic H$\beta$ flux.
This allows us to propagate the uncertainty in $A_V$ consistently into the metallicity estimates, rather than relying on a prior correction that is often uncertain in practice (detailed at the end of this subsection). 
To include a small number of low-mass, metal-poor galaxies, we modestly extrapolate the nominal validity range of the \Bian calibration from $\oh=7.8$ to 7.6.        
This extrapolation affects only a minor fraction of the sample: 12 out of 183 galaxies (6.6\%) have median metallicities below $\oh=7.8$, and only one stacked bin(Group 31, see Sect.~\ref{subsec:stack}), lies slightly below this boundary.
We further tested that excluding this stack has a negligible impact on both the MZR and FMR results, indicating that the modest extrapolation does not materially affect our main conclusions. 


From the de-reddened \Hb flux $f_{\Hb}$, we estimate the instantaneous SFR of our sample galaxies, from the Balmer lines luminosity. 
Assuming the \citet{Kennicutt:1998ki} calibration and the intrinsic Balmer decrement ratio of $\Ha/\Hb=2.86$ from case B recombination with $T_{\rm e}\sim10,000 {\rm K}$ for typical \HII regions, we calculate:
\begin{equation}
    \mathrm{SFR} = 4.65\times10^{-42} \frac{L(\Hb)}{\mathrm{[erg\,s^{-1}]}}\times 2.86\,[\Msun\,\mathrm{yr}^{-1}], 
\label{eq:sfr}
\end{equation}
suitable for the \citet{Chabrier:2003ki} initial mass function.
\rp{
The intrinsic Balmer decrement is only weakly sensitive to the assumed electron temperature \citep[e.g.,][]{Sandles:2024A&A}.
For typical HII-region conditions, varying $T_{\rm e}$ from $\sim8000$ to $\sim20000$ K changes $\mathrm{H}\alpha/\mathrm{H}\beta$ from approximately $\sim2.9$ to $\sim2.75$, corresponding to a change of $\lesssim4\%$ ($\lesssim0.02$ dex) in the H$\beta$-based SFR \citep[][table 4.4]{Osterbrock:2006agna}, which is small compared with the other SFR uncertainties.
}
This approach provides a valuable proxy of the ongoing star formation on a time scale of $\sim10$Myr, highly relevant for galaxies displaying strong nebular emission lines.
While the SFR derived from SED fitting (discussed later in Sect.~\ref{subsec:mass}) typically reflects a timescale of $\sim$100 Myr, adjusting it to 5 Myr shows good agreement with the H$\beta$-based SFR (discussed later in Sect.~\ref{subsec:sfrerr}).

Another advantage of our Bayesian joint-fitting method is that it treats $A_V$ as a nuisance parameter and marginalizes over its allowed range.
Thus, when $A_V$ is weakly constrained, the corresponding uncertainty is propagated into the posterior distribution of $12+\log(\mathrm{O/H})$, rather than imposed through a fixed external correction.
\rp{
The impact of dust attenuation depends strongly on the wavelength separation of the line ratio.
For diagnostics based on closely spaced emission lines, such as [OIII]/H$\beta$ (and also [SII]/H$\alpha$), the differential extinction is small.
Using the adopted \citet{Cardelli:1989ApJ} extinction law with $R_V=3.1$, a change of $\Delta A_V=1$ mag modifies $\log([\mathrm{OIII}]/\mathrm{H}\beta)$ by only $\sim0.02$ dex, corresponding to a metallicity shift of only $\sim0.03$ dex (around $12+\log(\mathrm{O/H})\simeq8.2$ when using the \Bian O3 calibration).
By contrast, for ratios involving more widely separated wavelengths, such as [OII]/H$\beta$, the effect is not negligible.
The same change of $\Delta A_V=1$ mag modifies the ratio in logarithm by $\sim0.15$ dex, which can translate into a metallicity shift of order $\sim0.1$ dex.
An inaccurate dust correction could therefore bias the inferred metallicity, especially when the metallicity estimate relies strongly on such long-baseline ratios.
For this reason, our Bayesian framework jointly constrains $A_V$, $f_{\mathrm{H}\beta}$, and $12+\log(\mathrm{O/H})$, so that the allowed range of dust attenuation is marginalized over and propagated into the metallicity posterior.
Therefore, even when $A_V$ is weakly constrained or has a broad posterior, its effect is not ignored but is reflected in the uncertainty of the inferred metallicity.
}

This treatment is particularly relevant for galaxies at $z\sim3$, where H$\alpha$ is not covered and the dust constraint relies mainly on the weaker H$\gamma$/H$\beta$ ratio.
One solution suggested by \citet{Sanders:2021ga} was to calibrate $A_V^\mathrm{gas}$ at $z\sim2$ using SED-derived $A_V^\mathrm{star}$ (Sect.~\ref{subsec:mass}) and then extrapolate this relation to $z\sim3$, while emphasizing that their calibration is specific to their sample.
As a sanity check, we also find that a similar calibration based on $A_V^\mathrm{star}$ is broadly consistent with the $A_V^\mathrm{gas}$ inferred by our method.
Nevertheless, caution is warranted because the correlation between $A_V^\mathrm{gas}$ derived from the Balmer decrement and $A_V^\mathrm{star}$ derived from SED fitting is neither tight nor universal, as widely reported in the literature \citep[e.g.,][]{Reddy:2015ho,Wang:2016um,Wang:2020bp}.
Furthermore, both the shape of the extinction curve and the typical attenuation in high-redshift galaxies remain debated \citep{Reddy:2020es,Salim:2020ia,Markov:2025NatAs}.
\rp{
We therefore also include [NeIII]$\lambda3869$/[OIII]$\lambda5007$ as an auxiliary constraint on dust attenuation, especially when H$\alpha$ is unavailable.
This is motivated by the nearly flat Ne3O3 relation first reported by \citet{Jones:2015ApJ...813..126J}, $\log([\mathrm{NeIII}]/[\mathrm{OIII}])\simeq -1.09$, and confirmed by recent DESIRED calibrations \citep{Rosales-Ortega:2026arXiv}, which show that \NeIII/\OIII is almost insensitive to metallicity and is not a viable metallicity diagnostic.
Although this flatness makes the ratio unsuitable for estimating metallicity, it allows \NeIII/\OIII to provide an auxiliary consistency check on dust attenuation, together with the Balmer-ratio information in our Bayesian framework.
}

\subsection{Stellar mass and AGN contamination}\label{subsec:mass}

In this section, we fit broad-band photometry to obtain the stellar mass $M_*$ of target galaxies. 
We directly use the combined photometric catalog released from the DJA as described in Sect.~\ref{sec:data}.

To estimate the stellar masses \Mstar of our sample galaxies, we use the \bagp software \citep{Carnall:2018gb} to fit the BC03 \citep{Bruzual:2003ck} models of SEDs to the photometric measurements derived above. 
As recommended, we assume the \citet{Chabrier:2003ki} initial mass function, a stellar metallicity range of $Z/Z_{\odot}\in(0, 2.5)$, 
the \citet{Calzetti:2000iy} extinction law with $A_V$ in the range of (0, 3). 
Stellar metallicity is a parameter in the Simple stellar-population (SSP) models used here only, as a by-product of the SED measurement of stellar mass $M_*$. 
We use the Double Power Law (DPL) model (other than the simple exponentially declining form) to capture the complex Star Formation History (SFH) of our galaxies at cosmic noon (rather than local universe) \citep{Reddy:2012hw,Pacifici:2015gg,Carnall:2019di}.
Further discussion of the impact of SFH is left to Sect.~\ref{subsec:sfh}.
The nebular emission component, including the lines and the continuum, is also added with the logarithm of the ionization parameter set to $\log_{10}U=-3$,
since our galaxies are exclusively strong line emitters by selection \citep{Byler:2017cn}.
The redshifts of our galaxies are fixed to their best-fit grism values, with a conservative uncertainty of $z_{\sigma} = 0.003$.
As a sanity check, we also use \textsc{Beagle} \citep{Chevallard:2016kc} to perform the SED fitting under similar settings, and the measured stellar mass $M_*$ is consistent with that using \textsc{Bagpipes}.

The metallicity diagnostics used in this work are strictly for star-forming regions/galaxies.
The interpretation of the line flux would be incorrect if the emission were powered by Active Galactic Nuclei (AGN), and \oh would be higher for AGN-host galaxies \citep{Li:2024MNRAS}.
However, the most traditional method of identifying AGN using the BPT diagram \citep{Baldwin:1981ev}, relies on $\NII~\lambda6584$, which is entirely blended with \Ha, as are other diagnostic lines at the spectral resolution of JWST/NIRISS slitless spectroscopy ($R\sim150$).
Fortunately, \citet{Juneau:2014ca} proposed an effective approach coined the mass-excitation (MEx) diagram, using \Mstar as a proxy for \NII/\Ha, which functions well at $z\sim0$ (\ie SDSS DR7).
\citet{Coil:2015dp} further modified the MEx demarcation by horizontally shifting these curves to high-\Mstar by 0.75 dex, which is shown to be more applicable to the MOSDEF sample \citep{Sanders:2021ga} at $z\sim2.3$. 
\rp{
Similar MEx-based AGN screening has also been adopted in recent JWST spectroscopic studies of high-redshift galaxies, where the \NII/\Ha-based BPT diagnostics are often unavailable or incomplete \citep[e.g.,][]{Chakraborty:2025ApJ, Jain:2026ApJ}.
}
We thus rely on this modified MEx, and prune 3 AGN-like sources from our galaxy sample.

\subsection{Stacking spectra} \label{subsec:stack}

\begin{figure*}
    \centering
    \includegraphics[width=1.\textwidth]{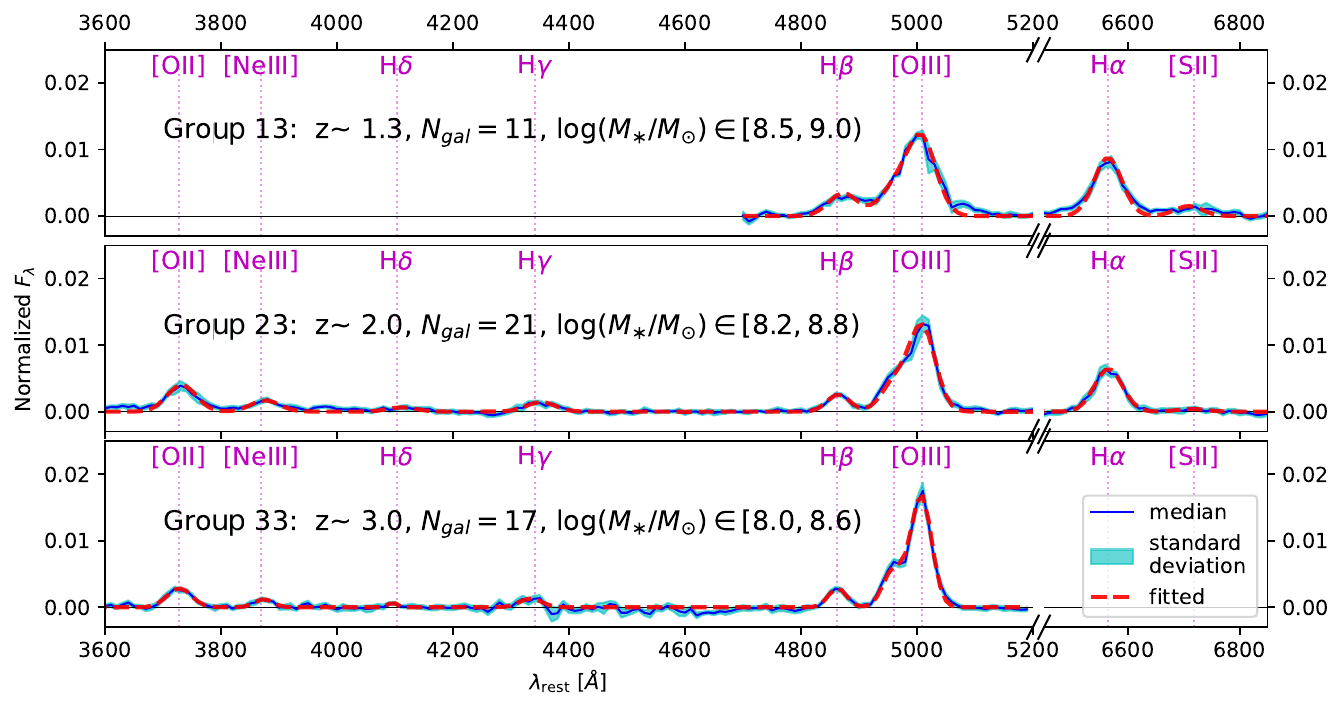}
    \caption{
    Three examples of stacked grism spectra for galaxies residing in similar mass bins at 3 redshift ranges (from top to bottom: $z\in[1.10,1.54],  [1.76,2.32], [2.61,3.43]$, respectively).
    In each panel, we mark the group ID (as defined in Tab.~\ref{tab:stack}), redshift range, number of galaxies $N_{gal}$, and corresponding mass range.
    For each set of spectra, the blue curves represent the median stacked spectrum, the cyan bands mark the bootstrapped flux uncertainties, and the red dashed curves show the best-fit Gaussian fits to multiple emission lines, while \SII, \Ha are across a discontinuous range among other lines (\ie, the \OIII$\lambda\lambda$4960,5008 doublets, \Hb, \Hg, \Hd, \NeIII, and \OII) in the broken axes at right parts.
    }
    \label{fig:stack}
\end{figure*}

Given the relatively low signal-to-noise ratio of our sample, we carry out an analysis of stacked spectra.
To this aim, our 183 spectroscopically confirmed galaxies in HUDF are divided into 3 redshift bins, depending on which of the 3 filters F115W, F150W, F200W the \OIII and \Hb fall into, and several mass bins, to ensure a reasonable number of galaxies in each bin, as presented in Tab.~\ref{tab:stack}. 
\rp{
We tested the sensitivity of our results to the adopted mass binning by repeating the stacking analysis with four alternative mass-binning schemes.
These tests were constructed by modestly shifting, widening, or narrowing the boundaries of the central three mass bins relative to the fiducial binning in Tab.~\ref{tab:stack}.
The resulting MZR parameters change only weakly, with maximum differences of $\Delta\beta \lesssim 0.1$ in slope and $\Delta Z^8 \lesssim 0.07$ dex in normalization, smaller than or comparable to the statistical uncertainties.
Thus, our conclusions are not driven by the exact choice of mass-bin boundaries.
}

Then we adopt the following stacking procedure, similar to that utilized by \citet{Henry:2021ju,Wang:2022ApJ...926...70W}:

\begin{enumerate}
    \item Subtract continuum models from the extracted grism spectra. The continua are constructed by \textsc{Grizli} combining two orients. 
    \item Normalize the continuum-subtracted spectrum of each object using its measured \OIII flux, to avoid excessive weighting toward objects with stronger line fluxes.
    \item De-redshift each normalized spectrum to its rest frame, and resample on the same wavelength grid using \textsc{SpectRes}\footnote{\url{https://spectres.readthedocs.io/en/latest/}} \citep{Carnall:2017arXiv} with the integrated flux preservation.
    \item Take the median value of the normalized fluxes (regenerated with errors) at each wavelength grid as the stacked spectrum.
    \item Re-create the stacked spectra (step 4) 1000 times with the bootstrapping replacement 
    and adopt their standard deviation as the uncertainty of the stacked spectrum.
\end{enumerate}

From the stacked spectra, we measure the emission line fluxes (\OII, \NeIII, \Hd, \Hg, \Hb, \OIII, \Ha, and \SII, if available, and normalized by \OIII) by fitting a set of Gaussian profiles.
The amplitude ratio of $\OIII~\lambda\lambda4960,5008$ doublets is fixed to 1:2.98 following \citet{Storey:2000jd}.  
We use the package \textsc{LMFit}\footnote{\url{https://lmfit.github.io/lmfit-py/}} to perform a nonlinear least-squares
minimization as shown in Fig.~\ref{fig:stack}, with the measured quantities summarized in Tab.~\ref{tab:stack}.
The stacked metallicity is estimated using the same methods as the individual galaxies outlined in Sect.~\ref{subsec:metal}.
Our later discussion will mainly focus on the stacked results. 

\rp{
Because the spectra are normalized by the observed \OIII flux before stacking, internal variations in dust attenuation could in principle affect some line ratios. 
However, \citet[][their Appendix B]{Henry:2021ju} explicitly compared dust correction before and after stacking in a similar grism-based framework and found negligible differences in the resulting stacked metallicities, suggesting that this effect is likely subdominant here.
Our methodological goal in Sect.~\ref{subsec:metal}, therefore, is to jointly constrain $A_v$ and metallicity within a single Bayesian framework, so that the uncertainty in dust attenuation is propagated into the posterior of 12+log(O/H), rather than imposed through a fixed external correction.
}

\begin{figure*}
    \centering
    \includegraphics[width=0.8\textwidth]{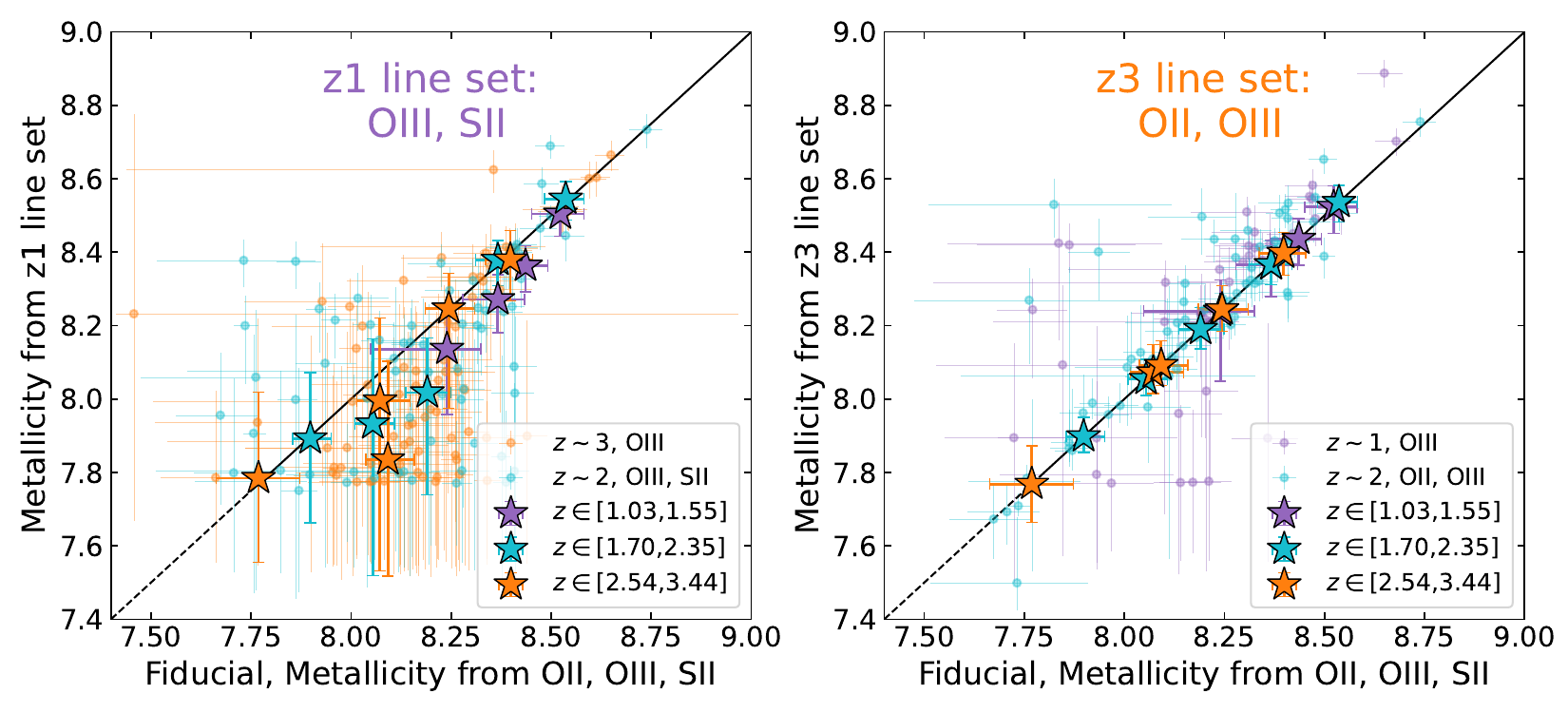}
    \caption{
Metallicity estimates derived from different combinations of emission lines.  
\textit{Left:} Line set at $z \sim 1$ using [OIII] and [SII] diagnostics.  
\textit{Right:} Line set at $z \sim 3$ using [OII] and [OIII] diagnostics.  
As a baseline, the abscissa in both panels employs the full set of [OII], [OIII], and [SII] metallicity diagnostics where available.  
Circles represent individual galaxies, while stars indicate stacked spectra.
    }
    \label{fig:lineset}    
\end{figure*}

It is important to assess whether the use of different emission-line sets across redshift bins introduces systematic biases in metallicity estimates.
As summarized in Tab.~\ref{tab:stack} and shown in Fig.~\ref{fig:stack}, our analysis employs \OII, \OIII, and \SII whenever available. 
However, the filter configuration (F115W, F150W, F200W) limits the coverage, such that \OII is absent at $z\sim1$ and \SII at $z\sim3$.
Our result in Fig.~\ref{fig:lineset} shows that the absence of \SII at $z\sim3$ has a negligible impact, as this line is intrinsically much fainter than \OII and \OIII. 
In contrast, the lack of \OII at $z\sim1$ may hinder the resolution of the well-known degeneracy in \OIII/H$\beta$ at low metallicities, potentially biasing the estimate downward. 
This effect is partially alleviated by the availability of \SII/H$\alpha$, and, more importantly, by the fact that the $z\sim1$ sample lies predominantly in the high-metallicity regime. 
Future observations with extended spectral coverage will be essential for minimizing such systematics and enabling more robust metallicity comparisons across redshift bins.

\begin{landscape}
\begin{table}       
  \scriptsize   
  \caption{Measured properties of the stacked spectra.}
  \label{tab:stack}
  \begin{tabular}{cccccccccccccccccc}
    \hline
    group & $\log(M_\ast/M_\odot)$ & $N_\mathrm{gal}$ & $M_\ast^{\rm med}$ & $z^{\rm med}$ & \mbox{[O\,\textsc{iii}]/H$\beta$} & \mbox{[O\,\textsc{ii}]/H$\beta$} & \mbox{[O\,\textsc{iii}]/[O\,\textsc{ii}]} & H$\gamma$/H$\beta$ & \mbox{[Ne\,\textsc{iii}]/[O\,\textsc{iii}]} & H$\alpha$/H$\beta$ & \mbox{[S\,\textsc{ii}]/H$\alpha$} & $A_\nu$ & \text{stack SFR} & \text{median SFR} & $12+\log(\mathrm{O/H})$ & $\log(\alpha\,\mu_{\rm gas})$ & $\log(\zeta_{\rm out})$ \\
    \hline
    \multicolumn{8}{c}{$1.10<z_\mathrm{grism}<1.54$} \\ 
11 & 6.3-8.0 & 9 & 7.82 & 1.41 & $6.68 \pm 0.35$ & ... & ... & ... & ... & $2.80 \pm 0.18$ & $0.07 \pm 0.01$ & $0.14_{-0.10}^{+0.19}$ &$0.77_{-0.07}^{+0.14}$ & 0.73 &$8.13_{-0.21}^{+0.10}$ &0.60 & $0.83_{-0.19}^{+0.32}$ \\
12 & 8.0-8.5 & 10 & 8.23 & 1.22 & $5.33 \pm 0.47$ & ... & ... & ... & ... & $2.54 \pm 0.25$ & $0.14 \pm 0.01$ & $0.15_{-0.11}^{+0.25}$ &$0.43_{-0.04}^{+0.10}$ & 0.92 &$8.27_{-0.10}^{+0.07}$ &0.46 & $0.67_{-0.13}^{+0.17}$ \\
13 & 8.5-9.0 & 11 & 8.67 & 1.31 & $4.60 \pm 0.36$ & ... & ... & ... & ... & $2.61 \pm 0.23$ & $0.18 \pm 0.03$ & $0.13_{-0.10}^{+0.22}$ &$0.44_{-0.04}^{+0.09}$ & 0.77 &$8.36_{-0.09}^{+0.06}$ &0.30 & $0.59_{-0.11}^{+0.14}$ \\
14 & 9.0-10.2 & 8 & 9.22 & 1.36 & $3.58 \pm 0.43$ & ... & ... & ... & ... & $3.15 \pm 0.42$ & $0.30 \pm 0.04$ & $0.46_{-0.30}^{+0.45}$ &$2.13_{-0.50}^{+0.99}$ & 2.27 &$8.50_{-0.07}^{+0.05}$ &0.11 & $0.43_{-0.10}^{+0.12}$ \\
    \hline
    \multicolumn{8}{c}{$1.76<z_\mathrm{grism}<2.32$} \\
21 & 6.3-7.6 & 9 & 7.11 & 1.97 & $8.75 \pm 1.14$ & $0.86 \pm 0.14$ & $10.23 \pm 1.06$ & $0.86 \pm 0.14$ & $0.09 \pm 0.01$ & $3.44 \pm 0.52$ & $0.07 \pm 0.03$ & $0.06_{-0.04}^{+0.10}$ &$0.78_{-0.06}^{+0.09}$ & 1.99 &$7.90_{-0.04}^{+0.05}$ &1.05 & $0.89_{-0.14}^{+0.10}$ \\
22 & 7.6-8.2 & 15 & 7.94 & 2.02 & $7.34 \pm 0.55$ & $1.19 \pm 0.11$ & $6.15 \pm 0.39$ & $0.42 \pm 0.07$ & $0.09 \pm 0.01$ & $3.54 \pm 0.32$ & $0.04 \pm 0.02$ & $0.29_{-0.17}^{+0.22}$ &$0.96_{-0.15}^{+0.23}$ & 2.41 &$8.05_{-0.05}^{+0.07}$ &0.76 & $0.86_{-0.14}^{+0.09}$ \\
23 & 8.2-8.8 & 21 & 8.64 & 2.07 & $7.60 \pm 0.70$ & $1.99 \pm 0.23$ & $3.82 \pm 0.30$ & $0.71 \pm 0.11$ & $0.11 \pm 0.01$ & $3.20 \pm 0.31$ & $0.07 \pm 0.02$ & $0.06_{-0.04}^{+0.09}$ &$1.31_{-0.07}^{+0.12}$ & 1.81 &$8.18_{-0.06}^{+0.05}$ &0.52 & $0.79_{-0.10}^{+0.09}$ \\
24 & 8.8-9.4 & 17 & 9.12 & 1.90 & $4.42 \pm 0.27$ & $2.14 \pm 0.17$ & $2.06 \pm 0.13$ & $0.26 \pm 0.04$ & $0.13 \pm 0.01$ & $2.57 \pm 0.18$ & $0.15 \pm 0.02$ & $0.08_{-0.06}^{+0.11}$ &$2.20_{-0.14}^{+0.24}$ & 2.87 &$8.36_{-0.05}^{+0.05}$ &0.35 & $0.56_{-0.10}^{+0.10}$ \\
25 & 9.4-10.2 & 8 & 9.82 & 2.03 & $3.11 \pm 0.27$ & $2.40 \pm 0.27$ & $1.29 \pm 0.11$ & $0.38 \pm 0.11$ & $0.19 \pm 0.02$ & $3.40 \pm 0.65$ & $0.16 \pm 0.07$ & $0.09_{-0.07}^{+0.15}$ &$3.57_{-0.35}^{+0.66}$ & 15.03 &$8.53_{-0.06}^{+0.05}$ &0.10 & $0.37_{-0.09}^{+0.10}$ \\
    \hline
    \multicolumn{8}{c}{$2.61<z_\mathrm{grism}<3.43$} \\
31 & 6.3-7.4 & 9 & 7.10 & 3.08 & $12.05 \pm 3.12$ & $0.49 \pm 0.20$ & $24.69 \pm 8.15$ & $2.12 \pm 0.67$ & $0.08 \pm 0.02$ & ... & ... & $0.57_{-0.44}^{+1.46}$ &$1.90_{-0.69}^{+6.98}$ & 3.85 &$7.78_{-0.11}^{+0.12}$ &1.15 & $1.05_{-0.37}^{+0.22}$ \\
32 & 7.4-8.0 & 23 & 7.72 & 2.98 & $7.68 \pm 0.91$ & $1.21 \pm 0.17$ & $6.36 \pm 0.56$ & $0.62 \pm 0.13$ & $0.07 \pm 0.01$ & ... & ... & $0.42_{-0.30}^{+0.96}$ &$3.18_{-0.93}^{+6.36}$ & 11.89 &$8.09_{-0.07}^{+0.12}$ &0.93 & $0.52_{-1.46}^{+0.22}$ \\
33 & 8.0-8.6 & 17 & 8.25 & 2.86 & $8.92 \pm 0.67$ & $1.33 \pm 0.13$ & $6.73 \pm 0.43$ & $0.56 \pm 0.12$ & $0.05 \pm 0.01$ & ... & ... & $0.64_{-0.40}^{+0.67}$ &$4.38_{-1.57}^{+4.90}$ & 11.46 &$8.11_{-0.06}^{+0.08}$ &0.75 & $0.76_{-0.19}^{+0.12}$ \\
34 & 8.6-9.2 & 18 & 8.80 & 2.97 & $6.44 \pm 0.58$ & $1.85 \pm 0.19$ & $3.49 \pm 0.22$ & $0.40 \pm 0.08$ & $0.07 \pm 0.01$ & ... & ... & $0.30_{-0.22}^{+0.44}$ &$3.57_{-0.79}^{+2.44}$ & 12.99 &$8.25_{-0.06}^{+0.07}$ &0.56 & $0.63_{-0.16}^{+0.12}$ \\
35 & 9.2-10.2 & 8 & 9.57 & 3.03 & $5.02 \pm 0.64$ & $2.42 \pm 0.32$ & $2.07 \pm 0.13$ & $0.39 \pm 0.13$ & $0.06 \pm 0.02$ & ... & ... & $0.29_{-0.21}^{+0.40}$ &$5.21_{-1.19}^{+3.39}$ & 5.90 &$8.40_{-0.06}^{+0.05}$ &0.29 & $0.52_{-0.11}^{+0.11}$ \\
\hline
  \end{tabular}
  \vspace{1ex}
  \textit{Notes.} The physics quantities derived from stacked spectra over 5 (or 4) mass intervals for 3 redshift ranges.
    The columns list in each bin for the galaxies the: (1) group label named by redshift and mass region, (2) mass interval, (3) number of galaxies, (4) median galaxy stellar mass ($\log(M_{\ast}/M_\odot)$), (5) median galaxy redshift, (6-12) multiple emission line flux ratios measured from the stacked spectra (while we use '...' to indicate the uncovered line) in Sect.~\ref{subsec:stack} (before dust correction), Bayesian inferred (13) dust extinction, (14) stack star formation rate (SFR), and (16) gas-phase metallicity in Sect.~\ref{subsec:metal}, (15) median galaxy SFR, (17-18) gas fraction and the metal loading factor of the outflowing galactic winds in Sect.~\ref{subsec:mzr}.
    The mass range and the median stellar mass $\log M_*^\mathrm{med}$ are both logarithmic values $\log (M_*/M_\odot)$.
    Since the stack spectrum normalized to the \OIII flux of each individual galaxy, the Bayesian inferred de-reddened \Hb flux is actually the ratio (column 6) relative to \OIII flux.
    Therefore the stack SFR (column 14) is derived from the relative \Hb flux multiplied by the median \OIII flux in each group.  
\end{table}  
\end{landscape}

\section{Results and Discussions}\label{sec:result}

From the first-epoch observations of the NGDEEP survey with JWST/NIRISS, we obtain a preliminary insight of the MZR for the most extensive current sample of dwarf galaxies at cosmic noon, spanning a stellar mass range of $M_*\in(10^{6.3},10^{10.2})\,M_{\odot}$ and redshifts $z\in(1.1,3.4)$. 
The large sample size further allows us to examine the potential existence of a FMR. 
In addition, we investigate in detail, for the first time, the systematic uncertainties in stellar mass estimates arising from different SFH assumptions in SED fitting, and evaluate their impact on MZR measurements.



\subsection{The Star-Forming Main Sequence}
\label{subsec:sfr}

\begin{figure}
    \centering
    \includegraphics[width=\columnwidth]{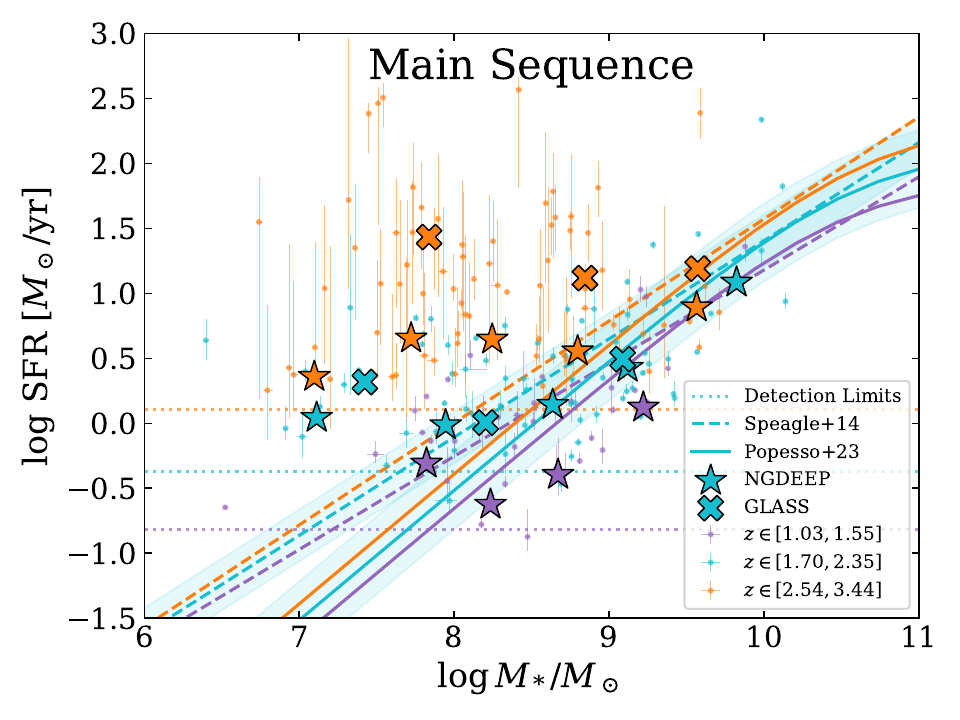}
    \caption{SFR-$M_*$ relation for our galaxy sample.
    The sample is divided into 3 redshift bins (according to their \OIII, \Hb detection in 3 filters of F115W, F150W, F200W), color-coded by purple, cyan, and orange, respectively.
    The stars mark the median value after mass binning.
    Three horizontal dotted lines marked the $3\sigma$ \Hb flux limit for each filter at each redshift (detailed in  Appendix~\ref{sec:appendix}), below which there is barely any \Hb detection.
    As a comparison, we also show the star-forming main sequence extrapolated from \citet[][dashed]{Speagle:2014dd} with $\pm0.2$ dex scatters, and from \citet[][solid]{Popesso:2023MNRAS} with $\pm0.3$ dex scatters. 
    The crosses label our previous results   \citep{He:2024ApJL} at the same region.
    }
    \label{fig:sfr}
\end{figure}

Before turning to the MZR and FMR, it is useful first to examine the SFR–$M_*$ relation of our sample, since it provides the baseline context for star-forming galaxies \citep{Gburek:2023ApJ}.
In Fig.~\ref{fig:sfr}, we can observe a clear tendency for the SFR to increase with redshift and to gradually flatten toward lower mass, 
\rp{broadly consistent with our previous GLASS-JWST results in the same region \citep{He:2024ApJL}. 
For reference, we compare our sample with the widely used SFMS parametrization of \citet{Speagle:2014dd}, which was originally constrained at $\log(M_*/M_\odot)\gtrsim9.7$, and with the more recent compilation-based SFMS of \citet{Popesso:2023MNRAS}, which extends the range down to $\log(M_*/M_\odot)\simeq8.5$ over $0<z<6$. 
\citet{Sanders:2021ga} showed that \citet{Speagle:2014dd} remains broadly consistent with their $z\sim2.3$ sample down to $\log(M_*/M_\odot)\simeq9$, while \citet{Merida:2023ApJ} tested its extrapolation to $\log(M_*/M_\odot)\simeq8.5$. 
Here we therefore use both of them only as reference MS loci, noting that their application below $\log(M_*/M_\odot)\simeq8.5$ is an extrapolation.
Our galaxies generally cluster around the MS at higher $M_*$ and lower redshift.
At lower $M_*$ and higher redshift, however, high-SFR galaxies dominate the detected sample, especially at $z\sim3$ and $M_*/M_\odot\lesssim3\times10^8$.
If the extrapolated low-mass MS remains approximately valid down to $M_*\sim10^7M_\odot$, and if the anti-correlation between SFR and O/H expected from an FMR-like dependence (detailed later in Sect.~\ref{subsec:fmr}) still applies at these masses, then this high-SFR selection bias ($\Delta\log\mathrm{SFR}\sim0.7-1.5$) could lead to an underestimate of $12+\log(\mathrm{O/H})$ at the low-mass end by up to $\sim0.1$ and $\sim0.2$ dex at $z\sim2$ and $z\sim3$, respectively, assuming ${\rm d(O/H)}/{\rm d(SFR)}=\beta\times a\simeq0.3\times0.5$.
If this is the case, the MZR of the underlying dwarf-galaxy population is likely to be flatter than the observed relation, possibly approaching a low-mass turnover or downward saturation, as also suggested by recent results at higher redshifts \citep{Curti:2023arXiv230408516C,Nakajima:2023arXiv230112825N}.
}

Currently, our understanding of the star-forming main sequence (SFMS) at masses down to $10^7 M_\odot$ at $z > 1$ remains very limited, especially when compared to the relatively well-studied regime above $10^9 M_\odot$ \citep[e.g.,][]{Popesso:2023MNRAS, Koprowski:2024A&A}.
Despite the pioneering efforts of \citet{Merida:2023ApJ} to extend the MS to dwarf galaxies using a large sample, completeness is only achieved for stellar masses $M_*/M_\odot \gtrsim 10^{8.0}, 10^{8.3}, 10^{8.6}$ at $z \sim 1, 2, 3$, respectively.  
Below their mass-representative limits, the majority (60\%) of their galaxies are located above the MS by approximately 0.7 dex, assuming the slope remains unchanged, thereby qualifying them as starburst galaxies. 
This trend is quite similar to that seen in our sample, where completeness is directly constrained by the observational limits shown in Fig.~\ref{fig:sfr}, even under these deep NIRISS observations.  
\citet{Merida:2023ApJ} also report the possibility of the MS slope flattening if mass completeness limits could extend below $10^8 M_\odot$. 
This suggests that populations of low-SFR quenching dwarf galaxies and dusty dwarf galaxies may contribute only minimally to the overall galaxy population \citep{Bisigello:2023A&A}.  
In summary, current studies of MZR and the FMR for low-mass galaxies at high redshift might be significantly affected by sample completeness \citep{Curti:2023arXiv230408516C,Li:2023ApJL,Revalski:2024arXiv,Stephenson:2024MNRAS,Stanton:2024arXiv,Sarkar2024arXiv240807974S}.
Deeper and larger datasets are essential to fully understand the SFMS and its slope evolution at masses as low as $10^7 M_\odot$ for $z > 1$.  

At the low-mass end, comparisons with the SFMS may be significantly
affected by sample completeness and by the selection of strong emission-line galaxies.


\begin{figure*} 
    \centering
    \includegraphics[width=\textwidth]{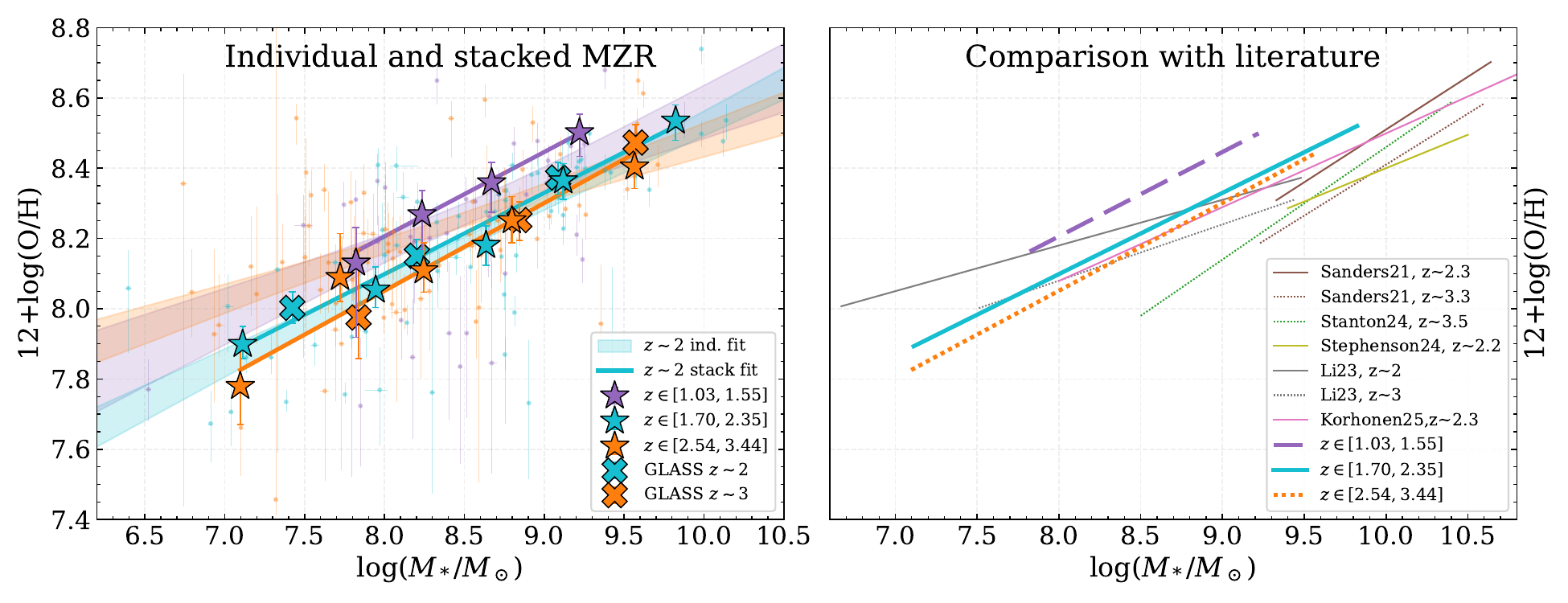}
    \caption{
    \textit{Left:}
    \rp{    
    MZR measurements for the combined sample of star-forming field galaxies from NGDEEP (stars; this work) and GLASS-JWST (crosses; previous results from \citealt{He:2024ApJL}). 
    }
    All individual galaxies (small dots) are divided into 3 redshift bins (purple, cyan, and orange), and further grouped into stellar-mass bins to obtain stacked metallicity measurements (large stars in the corresponding colors) in order to obtain robust linear fitted MZR (solid lines).
    \textit{Right:} 
    Comparison between our stacked MZR measurements (colored lines corresponding to the left panel) and representative literature results \citep{Sanders:2021ga,Stanton:2024arXiv,Stephenson:2024MNRAS,Li:2023ApJL,KorhonenCuestas:2025ApJ} at similar redshifts $z\sim2$ (dashed) or $z\sim3$ (dotted).
    For consistency, we only include literature measurements derived using, or converted to, the \citet{Bian:2018km} metallicity calibration.
    }
    \label{fig:mzr}
\end{figure*}

\subsection{The MZR for dwarf galaxies}
\label{subsec:mzr}

A clear positive correlation between gas-phase metallicity \oh, and stellar mass $M_*$ is found for both individual NGDEEP galaxies and stacked spectra, as shown in the left panel of Fig.~\ref{fig:mzr}.
This monotonic correlation is statistically confirmed using the Spearman correlation coefficient measured for individual galaxies in the three redshift ranges $z\in[1.10,1.54]$, $[1.76,2.32]$, and $[2.61,3.43]$.
The corresponding correlation coefficients are $0.463$, $0.728$, and $0.502$, with $p$-values of $3.42\times10^{-3}$, $8.79\times10^{-13}$, and $4.44\times10^{-6}$, respectively, all below the significance level of $0.05$.
\rp{
The MZR is characterized by a simple linear fit to the stacked spectra results:
\begin{equation}
    12+\log(\mathrm{O/H})
    = \beta \times \log(M_*/10^8M_\odot) + Z^8,
    \label{eq:mzr_def}
\end{equation}
where $\beta$ is the slope and $Z^8$ is the intercept at $M_* = 10^8 M_\odot$.
The detailed best-fitting parameters are listed in Tab.~\ref{tab:mzr_calib}.
}
For the fiducial \Bian\ calibration, the NGDEEP-only sample gives a nearly constant MZR slope of $\beta\simeq0.23$ across the three redshift bins, with median redshifts $z_\mathrm{med}=1.304$, $2.019$, and $2.935$.
At the same time, the intercept evolves monotonically, with the most prominent change occurring from $z\sim2$ to $z\sim1.3$.

\begin{table}
    \centering
    \caption{
    \rp{
    Comparison of the mass--metallicity relation (MZR) inferred from stacked spectra using different metallicity calibrations.
    The MZR is parameterized as $12+\log(\mathrm{O/H}) = \beta \times \log(M_*/10^8\,M_\odot) + Z^8$, where $\beta$ is the slope and $Z^8$ is the intercept at $M_*=10^8\,M_\odot$.
    The fiducial \Bian\ results are shown separately for the separate and the combined sample. 
    We also derive the MZR for the combined NGDEEP+GLASS sample over both the $z\sim2$ and $z\sim3$ range, since the subsamples of them yield broadly consistent metallicities and MZR fits for nearly all calibrations.
    A visual comparison of these calibration-dependent MZR measurements is shown in Figs.~\ref{fig:mzr_calib} and \ref{fig:mzr_z23}.
    Here we also list, in its first column, the nominal metallicity validity ranges of the comparison calibrations shown in Fig.~\ref{fig:calib}.
    }
  }
    \label{tab:mzr_calib}
  \begin{tabular}{l c c c}  
    \hline
    Calibration & $z$ range & slope $\beta$ & intercept $Z^8$ \\
    \hline
  Default:  & 1 & $0.240 \pm 0.014$ & $8.206 \pm 0.013$ \\
  \Bian, & 2 & $0.235 \pm 0.017$ & $8.087 \pm 0.018$ \\
  NGDEEP only & 3 & $0.223 \pm 0.029$ & $8.065 \pm 0.030$ \\
\noalign{\smallskip}\hline\noalign{\smallskip}

 \citealt{He:2024ApJL} & 2 & $0.223 \pm 0.017$ & $8.123 \pm 0.012$ \\
 \Bian, GLASS only & 3 & $0.294 \pm 0.010$ & $8.008 \pm 0.013$ \\
\noalign{\smallskip}\hline\noalign{\smallskip}
   Default: & 2 & $0.231 \pm 0.019$ & $8.098 \pm 0.018$ \\
   \Bian,  & 3 & $0.250 \pm 0.031$ & $8.052 \pm 0.030$ \\
   NGDEEP + GLASS & 2--3 & $0.238 \pm 0.015$ & $8.079 \pm 0.014$ \\
\noalign{\smallskip}\hline\noalign{\smallskip}
\multirow{2}{*}{\Nakajima}
    & 1 & $0.236 \pm 0.041$ & $8.116 \pm 0.039$ \\
    & 2 & $0.215 \pm 0.023$ & $8.028 \pm 0.023$ \\
\multirow{2}{*}{[6.9,8.9]}
    & 3 & $0.242 \pm 0.073$ & $7.966 \pm 0.070$ \\
    & 2--3 & $0.229 \pm 0.021$ & $8.002 \pm 0.022$ \\
\noalign{\smallskip}\hline\noalign{\smallskip}
\multirow{2}{*}{\Cataldi}
    & 1 & $0.266 \pm 0.042$ & $8.204 \pm 0.039$ \\
    & 2 & $0.242 \pm 0.027$ & $8.123 \pm 0.025$ \\
\multirow{2}{*}{[7.0,8.6]}
    & 3 & $0.356 \pm 0.131$ & $7.957 \pm 0.122$ \\
    & 2--3 & $0.302 \pm 0.050$ & $8.044 \pm 0.047$ \\
\noalign{\smallskip}\hline\noalign{\smallskip}
\multirow{2}{*}{\Sanders}
    & 1 & $0.240 \pm 0.044$ & $8.196 \pm 0.042$ \\
    & 2 & $0.251 \pm 0.025$ & $8.042 \pm 0.026$ \\
\multirow{2}{*}{[7.3,8.6]}
    & 3 & $0.265 \pm 0.039$ & $8.007 \pm 0.042$ \\
    & 2--3 & $0.255 \pm 0.016$ & $8.029 \pm 0.017$ \\
\noalign{\smallskip}\hline\noalign{\smallskip}
\multirow{2}{*}{\Chakra}
    & 1 & $0.391 \pm 0.084$ & $8.241 \pm 0.083$ \\
    & 2 & $0.345 \pm 0.062$ & $8.117 \pm 0.061$ \\
\multirow{2}{*}{[7.2,8.4]}
    & 3 & $0.430 \pm 0.123$ & $8.103 \pm 0.122$ \\
    & 2--3 & $0.386 \pm 0.044$ & $8.119 \pm 0.046$ \\
    \hline
  \end{tabular}
\end{table}

The MZR measured from the NGDEEP galaxies is broadly consistent with our previous \glass-\jwst result \citep{He:2024ApJL}, 
\rp{
as summarized in Tab.~\ref{tab:mzr_calib}, with differences within the current statistical uncertainties and consistent with expected field-to-field and sample-size variations.
}
Because the NGDEEP sample is approximately three times larger and extends to lower stellar masses, we combine the NGDEEP and \glass-\jwst samples to obtain a more statistically robust MZR measurement at cosmic noon, comprising 233 dwarf galaxies across 16 stacked bins with median redshifts of $z_\mathrm{med}=2.022$ and $2.947$, respectively.

We adopt these combined results as the default MZR measurements for the subsequent analysis and discussion.
Notably, the MZRs at $z\sim2$ and $z\sim3$ are similar, with the intercepts being indistinguishable within the $1\sigma$ uncertainties.
For reference, we also derive the MZR for the combined $z=2-3$ sample, with $z_\mathrm{med}=2.572$ in Tab.~\ref{tab:mzr_calib}.
\rp{
This combined  $z=2-3$  fit provides a compact comparison baseline for the calibration-dependent tests discussed below in Sect.~\ref{subsec:calib} and Fig.~\ref{fig:mzr_z23}.
}

The similarity can be quantified using the Chow test \citep{chow1960}, a statistical $F$-test commonly applied in econometrics to assess whether the true coefficients in two ordinary least-squares (OLS) regressions on different datasets are equal. 
\rp{
This test is appropriate for our purpose because we aim to determine whether the $z\sim2$ and $z\sim3$ stacked MZR measurements require different linear relations in the $M_*-\oh$ plane, i.e. different slopes and/or intercepts. 
By contrast, non-parametric distribution tests such as the Kolmogorov--Smirnov or Anderson--Darling tests are designed to compare one-dimensional distributions; applying them directly to $\oh$ would ignore the stellar-mass dependence of the MZR, while applying them to residuals would require first adopting a reference fitted relation.
}
While the Chow test is widely used in econometrics \citep[e.g.,][]{Sun:2019arXiv} and included in standard textbooks \citep[][Sect.~7-4c]{wooldridge2019}, it has been less commonly applied in astrophysics \citep{Markovskii:2016ApJ,Hayashi:2017ApJ,Kurapati:2021MNRAS}.
We note that the standard Chow test is based on unweighted OLS and does not fully account for heteroscedastic measurement uncertainties; here we therefore use it only as a simple diagnostic of whether separate linear fits are statistically required.


\rp{
The Chow test does not indicate a statistically significant difference between the $z\sim2$ and $z\sim3$ stacked MZR fits.
In other words, the current stacked data do not require separate slopes or intercepts for the two redshift bins.
Quantitatively, we obtain\footnote{Implemented in Python using the package available at \url{https://pypi.org/project/chowtest/}} $F=1.629$ and $p=0.237$, so the null hypothesis that the two stacked samples share the same linear relation cannot be rejected at the $p\leq0.05$ significance level.
This result should not be interpreted as proof that the two relations are intrinsically identical, but rather as evidence that any difference is not statistically significant given the current sample size.
The stacked sample remains relatively small, with only 8/16 stacked points for the split/group-stack measurements, which limits the sensitivity of the test.
As a check, using only the NGDEEP sample and excluding GLASS leads to an even lower statistical significance, with $F=0.394$ and $p=0.690$, mainly because the number of stacked points decreases to 5/10.
By contrast, the difference between the $z\sim1$ and $z\sim2$ stacked MZR fits is statistically significant, with $F=14.793$ and $p=2.052\times10^{-3}$.
We apply the Chow test only to the stacked MZR measurements as a simple diagnostic, and not to the individual-galaxy sample, since the outliers and objects with large uncertainties can strongly affect the residual sum of squares entering the Chow statistic, making the result difficult to interpret.
}

As shown in the right panel of Fig.~\ref{fig:mzr}, our slope, $\beta\simeq0.24\pm0.03$, is broadly consistent with those measured for more massive galaxies at cosmic noon, e.g., $0.29\pm0.02$ from \citet{Sanders:2021ga}, $0.32\pm0.09$ from \citet{Stanton:2024arXiv}, 
\rp{
$0.28\pm0.01$ from \citet{Jain:2026ApJ}, $0.21\pm0.02$ from \citet{KorhonenCuestas:2025ApJ}, and $0.27\pm0.04$ from the direct-$T_{\rm e}$ AURORA sample of \citet{Khostovan:2025arXiv}.
This agreement over nearly four orders of magnitude in stellar mass, $M_*\sim10^7-10^{11}M_\odot$, suggests that the mass dependence of the physical processes regulating metal enrichment, such as gas accretion, star formation, and metal-loaded outflows, may remain broadly similar across the dwarf and more massive star-forming galaxy populations at cosmic noon.
}
While \citet{Li:2023ApJL} reported extending MZR to even lower masses, their sample size remains limited, and their mass estimates for galaxies with $M_*<10^7M_\odot$ at $z\sim2$ were primarily derived from photometric data based on seven HST bands.
The notable difference in the slope between their results ($0.14\pm0.04$) and ours is likely attributed to the updated calibration files used in our NIRISS data reduction, and the updated JWST photometry 
\citep{He:2024ApJL}. 
\rp{
Another possible reason could be sample selection, which has also been discussed in \citep{Henry:2021ju,Raptis:2025arXiv}.
As described in Sect.~\ref{sec:data}, we adopt an \Hb-based SNR selection rather than an \OIII-based one to reduce metallicity-dependent selection effects.
An \OIII SNR cut would preferentially select high-excitation, strong-\OIII systems with large \OIII/\Hb ratios, which in our strong-line calibrations generally correspond to lower metallicities.
It could therefore miss intrinsically weak-\OIII but relatively strong-\Hb galaxies at the high-metallicity end, biasing the inferred metallicities low and potentially flattening the MZR slope.
}
This could also explain the shallow slope $0.19\pm0.10$ from \citet{Stephenson:2024MNRAS}.

Nevertheless, although not by much, our slope $\sim0.24$ is slightly shallower than the previous $\sim0.30$.
This slight difference may be due to \citet{Bian:2018km}'s systematic overestimation when calibrating metallicities below 7.8, or it may suggest a slight shift in the feedback mechanism at the low mass end.
\citet{Guo:2016ApJ,Pharo:2023ApJ} measured $\beta\sim0.3\pm0.02$ at $z\sim 0.6$ down to $10^8M_\odot$, in high agreement with models incorporating SN energy-driven winds (with predicted $\beta\sim0.33$ by \citealt{Guo:2016ApJ}).
Our result indicates that the momentum-driven wind (with predicted $\beta\sim0.17$) may gain more importance down to $10^8M_\odot$, following the \citet{Dave:2012MNRAS.421...98D} model.
However, this seems inconsistent with the basic idea that the outflow scalings are momentum-driven at high mass, and are energy-driven at low masses \citep[e.g.,][]{Dave:2013MNRAS}, yielding steeper slope at low mass end in MZRs.
One possible explanation is that the widely used equilibrium model may not adequately describe low-mass galaxies at cosmic noon, as suggested by \citet{Guo:2016ApJ}. Instead, these galaxies may reside in a more non-equilibrium state, as suggested by findings at even higher redshifts \citep{Li:2025arXiva}.
Although similar shallower low-mass MZR slopes have been reported at $z\sim1-2$\citep{Henry:2021ju,Hirschauer:2022ApJ,Revalski:2024arXiv}, 
\rp{
recent JWST/NIRSpec observations also suggest a comparable trend at higher redshift. 
In particular, \citet{Curti:2023arXiv230408516C} found a shallow low-mass-end MZR slope of $\beta=0.17\pm0.03$ for galaxies at $3<z<10$, and argued that this behaviour is broadly consistent with models including momentum-driven SNe winds. 
}
Nevertheless, it is still premature to determine at which stellar mass and redshift the MZR begins to flatten, or whether it eventually approaches a metallicity floor.

\rp{
In the mass interval $M_*\in(10^9,10^{10})\,M_\odot$ that overlaps with \citet{Sanders:2021ga}, our stack-based metallicities are modestly higher by about $0.1$ dex.
This offset is comparable to the typical uncertainty budget associated with strong-line metallicity estimates.
Although our median redshifts differ slightly from those of \citet{Sanders:2021ga}, we do not attribute this offset primarily to this redshift difference.
Based on the redshift evolution reported by \citet{Sanders:2021ga}, $\mathrm{d}\log(\mathrm{O/H})/\mathrm{d}z=-0.11$, the expected change over $\Delta z\sim0.3$ is only $\sim0.03$ dex.
We therefore regard this as a modest zero-point difference between the measurements that may reflect a combination of sample selection, line-set coverage, dust treatment, and implementation details, rather than assigning it to a single dominant cause.
This offset does not affect our main conclusions regarding the low-mass MZR slope or the lack of compelling evidence for an FMR in the present sample.
}

As described in the Sect.~\ref{sec:intro}, the MZR is one of the most important observational scaling relations for testing chemical-evolution and gas-regulator models of galaxy formation.
Its slope and normalization can be used to indirectly constrain quantities that are otherwise difficult to measure, such as the gas fraction $\mu_{\rm gas}$, the outflow metal-loading factor $\zeta_{\rm out}$, and the efficiency with which galaxies retain or lose newly produced metals.
As an example, we follow the analytical framework adopted by \citet{Sanders:2021ga}, based on the model of \citet{Peeples:2011ew}, to examine which physical processes are most relevant for dwarf galaxies.

\subsection{From Observed MZR to Physical Insight}

\begin{figure}
    \centering
    \includegraphics[width=\columnwidth]{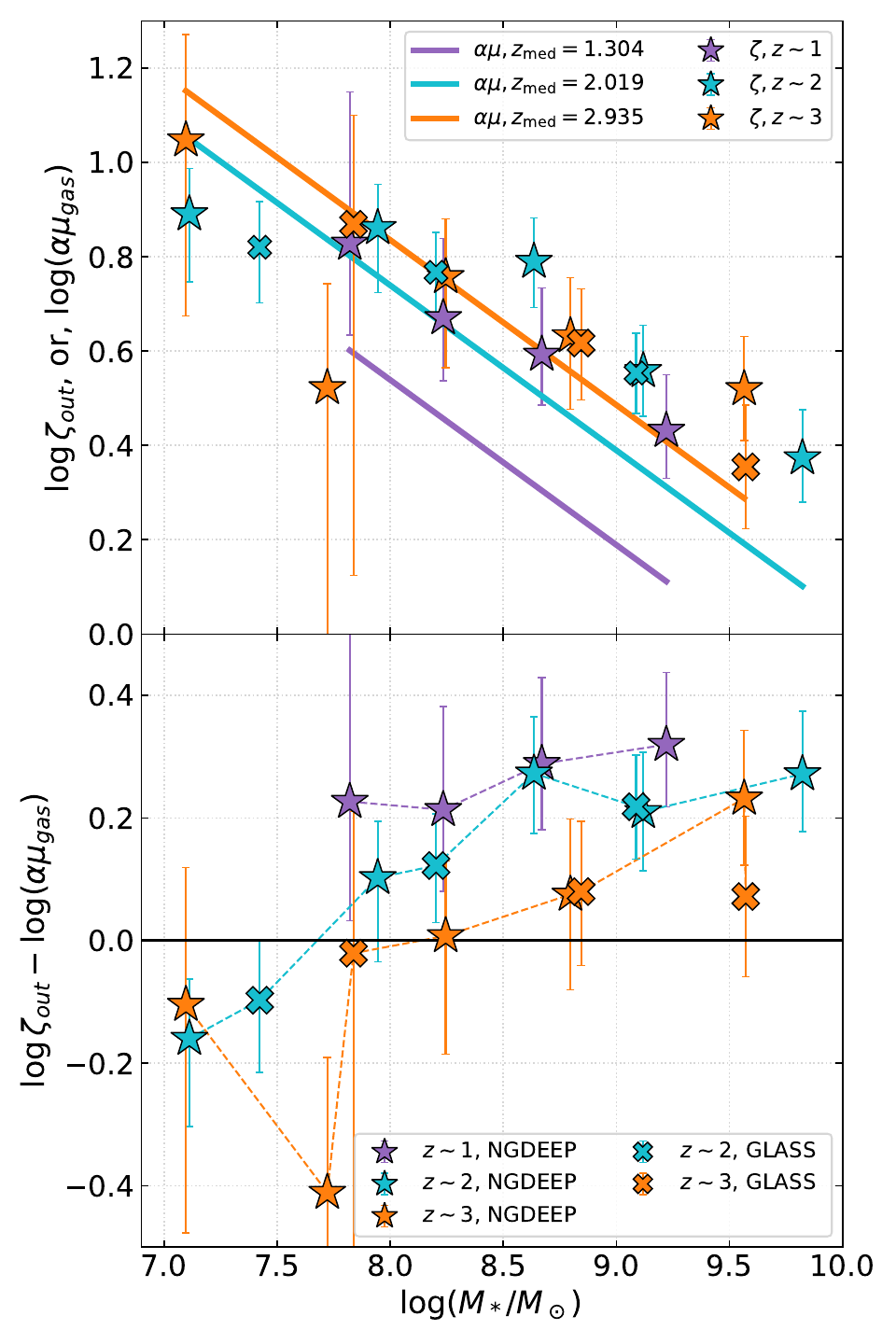}
    \caption{Outflow $\zeta_\mathrm{out}$ versus gas fraction $\alpha\cdot\mu_\mathrm{gas}$ as a function of $M_*$. 
    \textit{Top:} The solid lines show the gas fraction $\alpha\mu_\mathrm{gas}$ for each bin at their median redshift $z_\mathrm{med}$ \citep[quoted from][]{Tacconi:2018ApJ}. The filled stars (NGDEEP) and crosses (GLASS) show the calculated outflow $\zeta_\mathrm{out}$ following Eq.~\ref{eq:outflow}.
    All color and marker assignments here follow Fig.~\ref{fig:mzr}.  
    The lower bound of the orange star falling below the range is -0.9422.
    \textit{Bottom:} The difference between $\zeta_\mathrm{out}$ and $\alpha\mu_\mathrm{gas}$. 
    A positive value indicates that the scaling of outflow efficiency $\zeta_\mathrm{out}$ is more important for shaping the MZR than the scaling of gas fraction $\alpha\mu_\mathrm{gas}$ with $M_*$. 
    }   
    \label{fig:outflow}
\end{figure}

As described in the Sect.~\ref{sec:intro}, the MZR is one of the most important observational scaling relations 
\rp{
for testing chemical-evolution and gas-regulator models of galaxy formation.
Its slope and normalization can be used to indirectly constrain quantities that are otherwise difficult to measure, such as the gas fraction $\mu_{\rm gas}$, the outflow metal-loading factor $\zeta_{\rm out}$, and the efficiency with which galaxies retain or lose newly produced metals.
}
As an example, we follow the analytical framework adopted by \citet{Sanders:2021ga}, based on the model of \citet{Peeples:2011ew}, to examine which physical processes are most relevant for dwarf galaxies.
The (gas-phase) metallicity of the ISM: $Z_\mathrm{ISM}:= M_O/M_H$ (defined as the mass ratio, as opposed to the number density ratio log(O/H)) could be expressed as:
\begin{align}
    Z_\mathrm{ISM} &= \frac{y}{\zeta_\mathrm{out}-\zeta_\mathrm{in}+\alpha\cdot\mu_\mathrm{gas}+1}, \quad \text{where:} \label{eq:outflow} \\
    \zeta_\mathrm{out/in}:=\frac{Z_\mathrm{out/in}}{Z_\mathrm{ISM}} &\times\frac{\dot{M}_\mathrm{out/in}}{\mathrm{SFR}}, \alpha:=(1-R)\left(\frac{\mathrm{d}\log M_\mathrm{gas}}{\mathrm{d}\log M_\mathrm{*}}+ \frac{\mathrm{d}\log Z_\mathrm{ISM}}{\mathrm{d}\log M_\mathrm{*}}\right) \notag
\end{align}
They have consolidated the physical processes (their Eq.10 and Appx. C) into a set of key parameters in this formula to emphasize their interdependencies. 
These parameters include the metal-loading factor $\zeta_\mathrm{out/in}$ (a metallicity-specific counterpart to the mass-loading factor) that accounts for the effects of outflowing and inflowing galactic wind, the gas fraction $\mu_\mathrm{gas}$ and its scaling parameter $\alpha$, as well as the stellar metal yield $y$.
The effect of the gas fraction $\mu_\mathrm{gas}$ on $Z_\mathrm{ISM}$ is less intuitive compared to the effects of the inflow and outflow, but it can be understood as follows. 
A higher gas fraction implies a larger hydrogen reservoir $M_H$, which dilutes the metal content in the ISM. 
At the same time, a higher $\mu_\mathrm{gas}$ generally indicates that a smaller fraction of baryons has been converted into stars, leading to a lower cumulative metal production from stellar feedback. 
Together, these effects act to reduce $Z_\mathrm{ISM}$.
We follow the \citet{Tacconi:2018ApJ} best-fit relation: $\log\mu_\mathrm{gas}= 0.12-0.35(\log M_*-10.7) -3.62(\log(1+z)-0.66)^2$, and \citet{Sanders:2021ga} assumptions: the coefficient $\alpha=0.7\cdot(0.65+\beta)$ , the nucleosynthetic stellar yield for oxygen $ y\cong y_\mathrm{O}=0.015=Z_\mathrm{ISM}\cdot10^{9.2-(12+\log(\mathrm{O/H}))}$, and the metal loading factors of inflowing gas accretion $\zeta_\mathrm{in}=0$.
Our aim is to compute $\zeta_\mathrm{out}(M_*)$ based on $\mu_\mathrm{gas}$, performed at each stack-based point $(M_*, \mathrm{O/H})$ with the observed MZR slope $\beta$, as shown in Fig.~\ref{fig:outflow}.

We present the calculated $\zeta_\mathrm{out}$ (stars and crosses) and the quoted $\alpha\mu_\mathrm{gas}$ (solid lines) in the top panel. The bottom panel of Fig.~\ref{fig:outflow} displays their difference, allowing a comparison of their relative importance in determining the metallicity at different redshifts (indicated by colors).
We find that at $z\sim1.3$, $\zeta_\mathrm{out}\approx2\times \alpha\mu_\mathrm{gas}$ ($\sim0.3$ dex upper) and remain dominant down to $M_*\sim10^{7.5}M_\odot$.
At $z\sim2.0$, however, $\alpha\mu_\mathrm{gas}$ gradually overtakes $\zeta_\mathrm{out}$ as $M_*$ decreases to $\sim10^{7.7}M_\odot$. 
This transition occurs earlier, at $\log(M_*/M_\odot)\approx8.2$, for $z\approx3$, albeit not as significantly as for the higher-mass counterpart \citep{Sanders:2021ga}, especially when the typical uncertainty $\Delta\log\mu_\mathrm{gas}\simeq0.2$ is considered.

\rp{
This comparison suggests that, for high-redshift dwarf galaxies ($z\gtrsim2$, $M_*\lesssim10^8M_\odot$), the rapidly increasing gas content may become at least as important as the effective metal-loss term in setting the observed MZR normalization.
However, this should not be interpreted as evidence that pristine-gas dilution alone dominates the metallicity evolution of dwarf galaxies.
Recently, \citet{Laseter:2025arXiv} investigated the gas content and effective yields of low-mass galaxies using local \OIII$\lambda4363$ emitters and complementary high-redshift dwarf galaxies.
They found that these systems are generally gas rich, but that their large metallicity diversity cannot be explained by gas fraction alone.
Instead, their effective-yield analysis suggests that metallicity variations in low-mass, gas-rich systems are more closely connected to recent star formation and metal-enriched outflows than to pristine-gas dilution alone.
Our result is therefore complementary to this physical picture.
The increase of $\alpha\mu_\mathrm{gas}$ relative to $\zeta_\mathrm{out}$ in our $z\gtrsim2$ low-mass bins indicates that the gas reservoir is likely a key ingredient in shaping the low-mass MZR at cosmic noon.
At the same time, the instantaneous metal retention in dwarf galaxies may also be regulated by bursty star formation and enriched outflows, rather than by a simple competition between gas dilution and metal loss.
}

\rp{
\subsection{Systematic Uncertainties in Metallicity}
\label{subsec:calib}
}

Different calibrations can introduce systematic offsets in the absolute metallicity scale and affect the inferred MZR \citep{Kewley:2008be}.
Recent work has therefore emphasized the need to use calibrations appropriate for the target sample, or to explicitly quantify the calibration-induced systematic uncertainty \citep[e.g.][]{Scholte:2025MNRAS,Stanton:2024arXiv,Stephenson:2024MNRAS,KorhonenCuestas:2025ApJ,Raptis:2025arXiv}.
A systematic compilation and comparison of local, local-analog, and high-redshift strong-line calibrations is also presented by \citet{Rosales-Ortega:2026arXiv}.
One approach to minimizing calibration-related systematics (e.g., the EXCELS survey), is to construct or validate strong-line calibrations using a direct-$T_{\rm e}$ subsample that is matched to the target galaxy population \citep{Scholte:2025MNRAS}, and subsequently apply these internal calibrations to their full sample \citep{Stanton:2026MNRAS}.
In this section, we quantify the impact of calibration choice on our measurements.
While \Bian is adopted as the fiducial calibration throughout the main analysis, we repeat the metallicity inference and MZR fitting using four additional representative calibrations.
Our goal is not to identify a uniquely preferred calibration, but to assess whether the absolute metallicity scale and MZR parameters are robust to reasonable alternative calibrations.    

As introduced in Sect.~\ref{subsec:metal}, the calibrations used in this section for comparison are chosen to span local-analog \citep{Bian:2018km}, low-metallicity/high-excitation \citep{Nakajima:2022ApJS..262....3N}, cosmic-noon \citep{Cataldi:2025A&A}, and broad high-redshift empirical frameworks \citep{Sanders:2025arXiv,Chakraborty:2025ApJ}.
Our fiducial calibration is \Bian, which is based on local analogs selected to match the locus of $z\sim2$ star-forming galaxies on the BPT diagram.
Their calibrations are anchored to direct-$T_{\rm e}$ oxygen abundances measured from stacked spectra with detected [O~{\sc iii}]~$\lambda4363$, and cover the range of $\oh \in [7.8,8.4]$.
We adopt \Bian as our fiducial calibration because it is tailored to high-redshift-like ionization conditions and allows a direct comparison with previous $z\sim2-3$ studies using similar diagnostics \citep{He:2024ApJL}.
\rp{
This choice is supported by \citet{Gburek:2023ApJ}, who found that the \Bian local-analog calibration best reproduces the direct-method oxygen abundance of a $z\simeq2.3$ dwarf-galaxy composite at fixed strong-line ratio.
Nevertheless, recent JWST direct-$T_{\rm e}$ studies now provide high-redshift and cosmic-noon calibrations that can differ from local-analog relations because of differences in ionization conditions, excitation, and calibration-sample coverage (\Cataldi,\Sanders,\Chakra).
We therefore retain \Bian as the fiducial calibration, but explicitly quantify the calibration-dependent systematic uncertainty using the alternative calibrations described above.
}

As a complementary local empirical calibration, \citet[][hereafter, \Nakajima]{Nakajima:2022ApJS..262....3N} combines local SDSS star-forming galaxies with a large compilation of extremely metal-poor galaxies, and is fully anchored to direct-method metallicities over a broad range of $[6.9,8.9]$.
Compared with \Bian, \Nakajima provides a wider metallicity baseline and is particularly useful for testing the low-metallicity and high-excitation regime relevant to low-mass galaxies.
\citet[][hereafter \Cataldi]{Cataldi:2025A&A} present the MARTA calibration based on ultra-deep JWST/NIRSpec observations of 16 star-forming galaxies at $z\sim2-3$ with detections of one or more auroral lines. 
Ca25 provides the most directly redshift-matched cosmic-noon comparison.
\citet[][hereafter \Sanders]{Sanders:2025arXiv} provide a broader high-redshift framework from the AURORA survey, combining 41 new auroral-line galaxies with 98 literature galaxies at $z=1.3-10.6$.
Their calibrations cover a wide range of high-redshift galaxy conditions and serve as a broad modern high-redshift reference for our comparison.
Finally, \citet[][hereafter \Chakra]{Chakraborty:2025ApJ} derive empirical calibrations from 42 new JWST/NIRSpec galaxies at $z=3-10$, combined with 25 literature galaxies, yielding a total sample of 67 high-redshift galaxies, with direct-method metallicities spanning $\oh \in [7.2-8.4]$.
Since this sample is concentrated at a higher redshift, we use \Chakra mainly as an additional high-redshift/high-excitation comparison.
These four calibrations are therefore well suited for assessing whether our main results depend sensitively on the adopted strong-line calibrations.

\begin{figure*}
    \centering
    \includegraphics[width=0.8\textwidth]{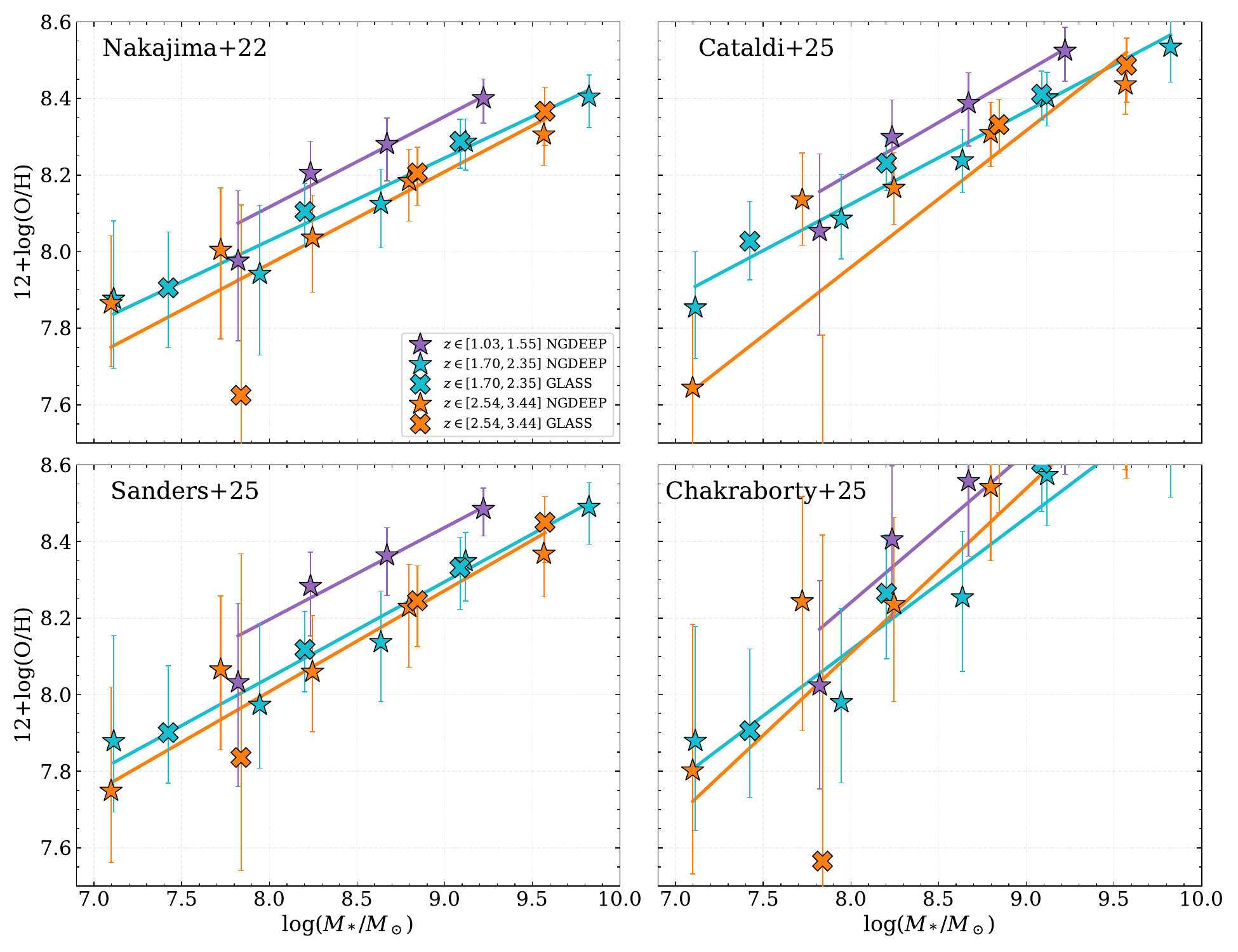}
    \caption{
    \rp{
    MZR measurements inferred from the $\mathrm{O}_3$--$\mathrm{O}_2$ diagnostics using different metallicity calibrations.
    The four panels correspond to \Nakajima,\Cataldi,\Sanders,\Chakra, respectively.
    All markers and line colours follow Fig.~\ref{fig:mzr}, and similar axis limits are retained to facilitate comparison with the fiducial \Bian result.
    Especially for the lower right \Chakra panel, metallicities above its upper validity range should be interpreted with caution.
    }
    }
    \label{fig:mzr_calib}
\end{figure*}

\begin{figure}
    \centering
    \includegraphics[width=\columnwidth]{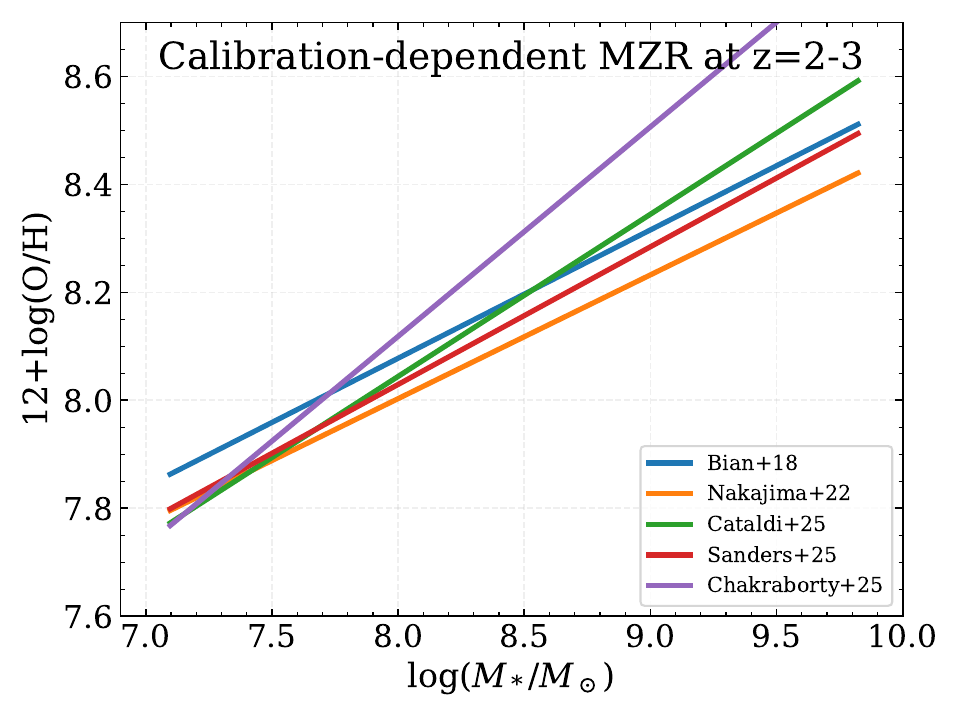}
    \caption{
    \rp{
    Comparison of the MZR inferred using the five metallicity calibrations adopted in this work.
    To avoid overcrowding the figure and to complement Fig.~\ref{fig:mzr_calib}, we show only the MZR measured from the combined $z=2-3$ sample for each calibration.
    This choice is motivated by the fact that the $z\sim2$ and $z\sim3$ subsamples yield broadly consistent metallicities and MZR fits for nearly all calibrations.
    The corresponding best-fitting parameters are summarized in Tab.~\ref{tab:mzr_calib}.
    }
    }
    \label{fig:mzr_z23}
\end{figure}


We repeat the MZR measurements while keeping all other steps unchanged, based on the same stacked spectra listed in Tab.~\ref{tab:stack}.
The resulting metallicities and qualitative MZR trend are stable under all calibrations considered here.
As shown in Fig.~\ref{fig:mzr_calib} with the corresponding best-fitting parameters summarized in Tab.~\ref{tab:mzr_calib}, the four alternative calibrations, \Nakajima, \Cataldi, \Sanders, and \Chakra, all preserve a clear positive correlation between $M_*$ and \oh.
This indicates that the existence of a positive MZR in our dwarf-galaxy sample is not driven by the specific choice of the fiducial \Bian\ calibration.
Quantitatively, the inferred slopes are also broadly consistent with the fiducial results discussed in Sect.~\ref{subsec:mzr}, remaining in the range $\beta\simeq0.23-0.27$.
The main exceptions are the $z\sim3$ result based on \Cataldi\ and the fits based on \Chakra, which yield steeper slopes and will be discussed further below.
The intercepts show the same overall evolutionary behaviour as in the fiducial analysis: the offset between $z\sim1$ and $z\sim2$ is larger than that between $z\sim2$ and $z\sim3$.
Moreover, for nearly all calibrations, the $z\sim2$ and $z\sim3$ MZR fits are consistent within the uncertainties, supporting the use of a combined $z=2-3$ sample for a compact calibration comparison, as shown in Fig.~\ref{fig:mzr_z23}.
This presentation avoids overcrowding Fig.~\ref{fig:mzr_calib} while providing a direct view of the calibration-dependent systematic shifts.
Different calibrations shift the absolute metallicity scale by up to $\Delta Z\sim0.5$ dex at fixed $M_*$, while the fitted slopes vary only modestly in most cases, with $\Delta\beta \lesssim 0.05$.
Therefore, the calibration choice mainly affects the MZR normalization and quantifies the systematic uncertainty of the strong-line metallicity scale \citep{Kewley:2008be}, but does not alter our main conclusion that low-mass galaxies at cosmic noon follow a positive MZR.


The relatively steep $z\sim3$ MZR obtained with the \Cataldi\ calibration is mainly driven by the two lowest-mass stacked bins, which have inferred metallicities of $\oh<7.7$.
Apart from these two bins, the stacked measurements follow a trend broadly similar to those obtained with the other calibrations.
This behaviour likely reflects the reduced constraining power of the O3 diagnostic near the low-metallicity turnover at around 7.8, where the relation becomes relatively flat and potentially degenerate.
In this regime, the metallicity inference is therefore more strongly driven by the O2 diagnostic.
However, the low-metallicity behaviour of O2 differs substantially among empirical calibrations in Fig.~\ref{fig:calib}, and is expected to be sensitive to the size and physical properties of the calibration sample, including the ionization state of the galaxies.
The \Nakajima\ calibration shows a qualitatively similar trend in O2, but in that case only one stack falls below $\oh\simeq7.7$, so the fitted MZR slope is less affected.
The other case is that \Chakra\ calibration produces systematically higher metallicities and a steeper MZR than the fiducial \Bian\ result.
This difference can be traced to the shape of the \Chakra\ O3 diagnostic shown in Fig.~\ref{fig:calib}, which departs from the other calibrations most strongly at $\oh\gtrsim8.2$.
As a result, high-metallicity stacks are shifted upward, leading to a larger MZR slope.
Given the limited calibration leverage at high metallicity and the fact that some of our points approach or exceed their nominal upper validity range, we regard the \Chakra\ result primarily as a stress test of high-redshift calibration systematics.


Overall, our tests show that the qualitative presence of a positive MZR is robust against the choice of metallicity calibration.
However, a quantitative interpretation of the MZR slope, intercept, and absolute metallicity scale requires an appropriate calibration.
Here, ''\textit{appropriate}'' does not simply mean that the calibration was derived at a similar redshift.
Rather, the calibration sample should span the relevant physical parameter space of the target galaxies, including metallicity, ionization state, excitation, density, and abundance patterns \citep{Rosales-Ortega:2026arXiv}.
Although cosmic-noon \Cataldi\ calibration provides the closest redshift match to our sample, its low-metallicity behaviour can still be sensitive to the limited leverage of the available auroral-line sample.
Conversely, local-analog \Bian and \Nakajima, can provide stable results when their calibration samples better match the excitation and ionization conditions of the galaxies under study.
Similarly, a broad compilation \Sanders can partly mitigate sample-selection effects by covering a wider range of high-redshift auroral-line measurements.
The \Chakra\ calibration, on the other hand, demonstrates that a high-redshift calibration should not be applied uncritically when the target galaxies approach or exceed the range where it is well constrained.
This interpretation is consistent with the conclusion of \citet{Rosales-Ortega:2026arXiv}, who argue that the key issue is not whether a calibration is local or high-redshift, but whether the calibrating sample adequately covers the multi-dimensional parameter space relevant to the target population.
In practice, this requirement is not always easy to verify, especially for low-resolution slitless spectra where the full set of classical diagnostic diagrams, such as the BPT diagram, may not be available.
Therefore, some empirical uncertainty in the calibration choice is unavoidable.
In such cases, applying several representative calibrations and explicitly quantifying the resulting spread provides a practical and transparent way to assess the systematic uncertainty in strong-line metallicity measurements.

\subsection{Systematic Uncertainties in Stellar Mass}
\label{subsec:sfh}


\begin{figure*}
  \centering
    \includegraphics[width=\textwidth]{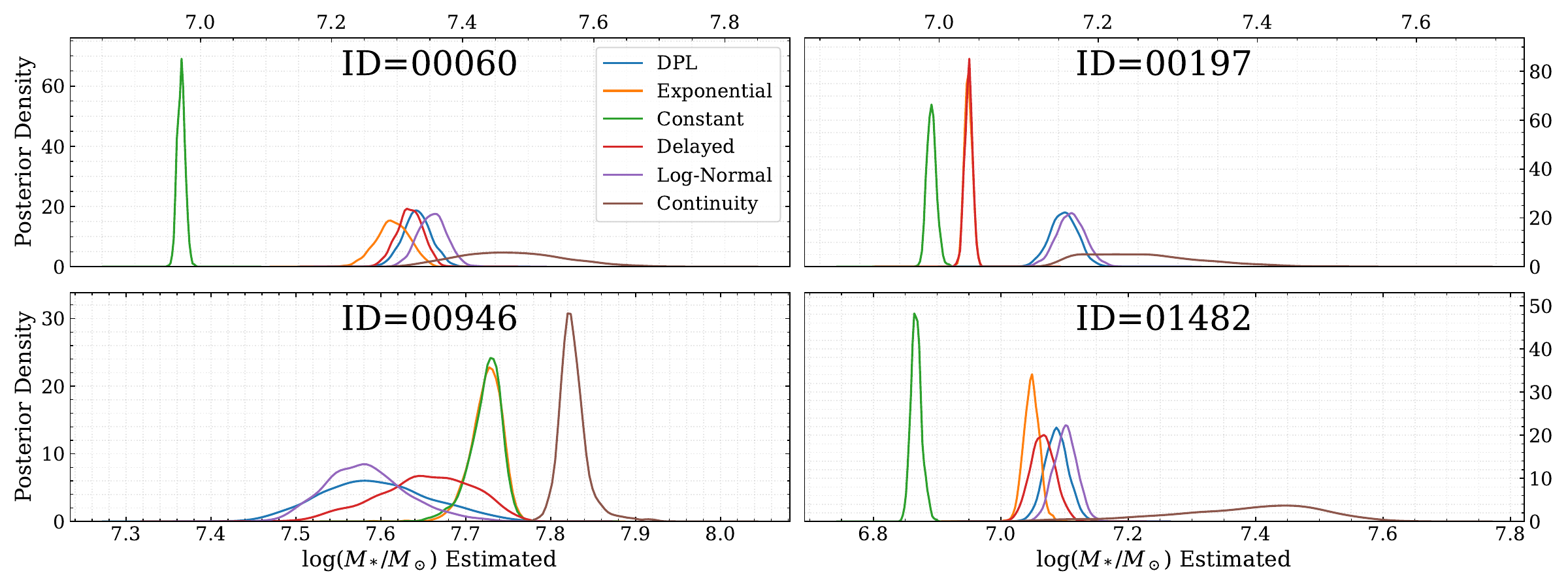}
    \caption{ Mass posterior distributions inferred under different SFH priors for 4 example dwarf galaxies.
    For comparison, five parametric models (Double Power Law, Exponential Decline, Constant, Delayed-$\tau$, and Lognormal) and one non-parametric model (Continuity) are used.
    Each distribution shows the KDE of the posterior, where broader posteriors have lower peak heights. 
    \rp{
    Most SFH models yield median stellar masses consistent at the $\sim0.1$ dex level, while some assumptions lead to larger offsets of $\sim0.2-0.3$ dex for individual galaxies.
    }
    This is broadly consistent with previous studies showing that SFH choices introduce systematic uncertainties in SED-derived stellar masses \citep{Carnall:2019di,Leja:2019hu}.
    }
    \label{fig:sfhpost}
\end{figure*}

\begin{figure*}
  \centering
    \includegraphics[width=\textwidth]{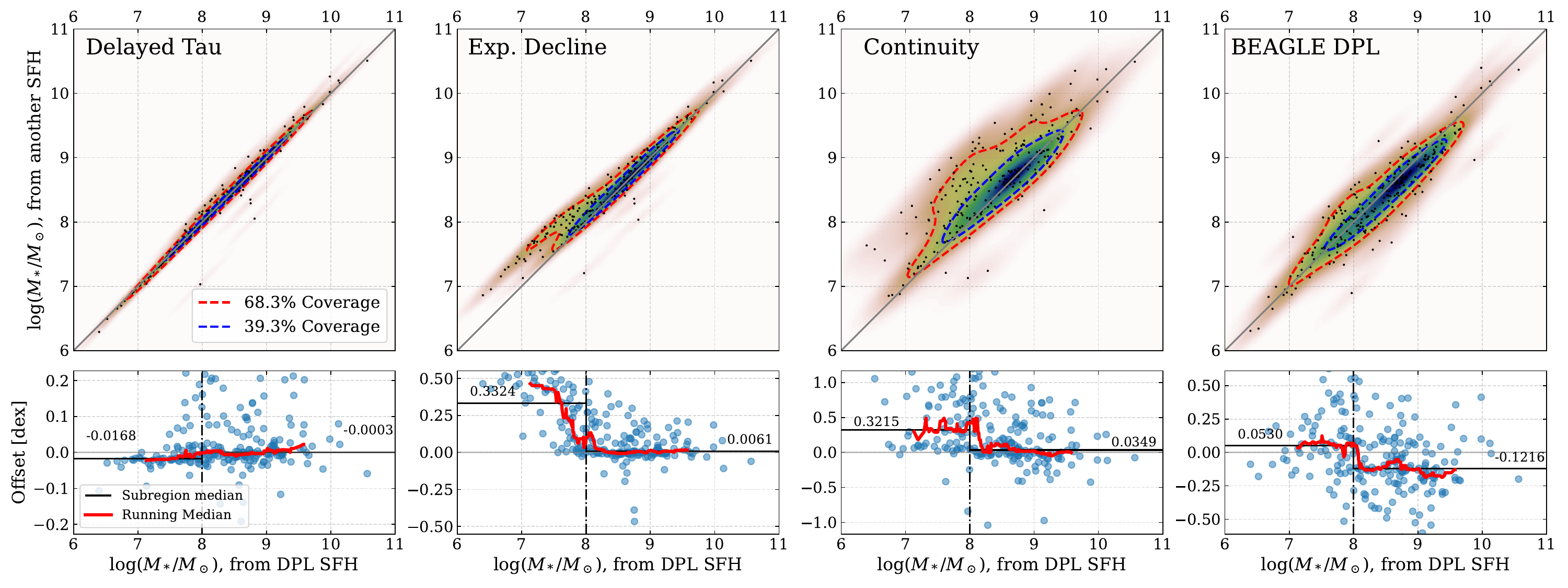}
    \caption{ 
    \textit{Top:} Comparison of inferred $M_*$ from 3 selected SFH models (Delayed-Tau, Exponential Decline, and Continuity), as well as the additional independent SED–fitting code \textsc{Beagle}, and the reference Double Power Law (DPL) model.
    The colored regions are the 2D KDE of the mass distribution (small dots), with red and blue contours enclosing the 68.3\% ($1\sigma$ in 1D) and 39.3\% ($1\sigma$ in 2D) of the volume, around the one-to-one line (gray lines).
    \textit{Bottom:} Stellar mass offset as a function of the reference mass.
    The red lines represent the running median of the offset (light blue circles).
    We also label the subregion median (black horizontal lines and the text) of the left and right subintervals separated by $\log(M_*/M_\odot)=8$. 
    Evidently, the dwarf galaxy masses estimated using different SFH models, can result in systematically higher uncertainties up to 0.33 dex, consistent with \citep{Lower:2020kc, delosReyes:2024arXiv}.
    }
    \label{fig:sfhoffset}
\end{figure*}

Given that the MZR relies directly on stellar mass estimates, in this section we explore the systematic uncertainties in stellar mass $M_*$ estimation arising from the selection of different SFH priors,  and in turn, assess their impact on the derived MZR.

The current standard approach to measuring $M_*$ relies on fitting multi-band photometry using spectral energy distribution (SED) fitting techniques, rather than the traditional mass-to-light ratio calibrations \citep{Stephenson:2024MNRAS}.
Significant discussions remain on how accurately $M_*$ can be recovered using existing SED fitting software (e.g., \textsc{Bagpipes} \citep{Carnall:2018gb}; \textsc{Beagle} \citep{Chevallard:2016kc}; \textsc{Prospector} \citep{Leja:2017dy}; \textsc{FAST} \citep{Kriek:2009cs}), either by using simulation data generated by the corresponding package \citep{Carnall:2019di, Leja:2019hu}, or by using mock galaxies from other cosmological simulations \citep{Lower:2020kc, delosReyes:2024arXiv}.
As one of the parameterizations of the SED introduced in Sect.~\ref{subsec:mass}, the choice of the SFH will directly impact the SFR and $M_*$ estimation, as the stellar mass is the integral of the SFH across time (after accounting for the stellar remnants fraction and mass loss). 
Three different families of SFH models are commonly used: parameterized, non-parametric, and models drawn directly from simulations. 
\rp{
The parameterized model is a description of the entire SFH using a single analytic function, such as an exponentially declining, delayed-$\tau$, lognormal, or double power-law form, and is therefore computationally efficient for large samples \citep{Carnall:2019di}.
The non-parametric model describes the SFH using multiple components or piecewise values over discrete time intervals, thus offering greater flexibility to fit diverse evolutionary scenarios, although the inferred SFH can depend sensitively on the adopted priors \citep{Leja:2019hu,Lower:2020kc}.
}
In studies concerned with MZR and FMR, the most common approach is parametric SFH model, e.g. Constant \citep{Sanders:2021ga,Li:2023ApJL, Stanton:2024arXiv}, Exponential decline \citep{Wang:2022ApJ...926...70W,Wang:2022ApJ...938L..16W,Revalski:2024arXiv}, Delayed-Tau \citep{Matthee:2023ApJ,Curti:2023arXiv230408516C}, while the non-parametric and customized models are less common but are gaining recognition \citep{Henry:2021ju,Nakajima:2023arXiv230112825N}.
It is now broadly agreed that stellar mass $M_*$ is the most robust among other parameters (e.g. Age and SFR) derived from SED, when using different SFH models \citep{Carnall:2019di,Leja:2019hu}, in the use of different SFH models.
However, when mock galaxies from hydrodynamical cosmological simulations are used, particularly those with rapid SFH variations, parameterized SFH models might introduce systematic error in mass estimation, both for normal galaxies \citep{Lower:2020kc} and dwarf galaxies \citep{delosReyes:2024arXiv}.



Following Sect.~\ref{subsec:mass}, we used the photometric catalog compiled by DJA as detailed in Sect.~\ref{sec:data}, to assess the systematic uncertainties of the stellar mass $M_*$ estimation introduced by different SFH priors.
Initially, we compared the $M_*$ posterior distributions for four sample dwarf galaxies using several popular SFH models in \textsc{Bagpipes}: five parameterized (Double Power Law, Exponential Decline, Constant, Delayed-Tau, and Lognormal) and one non-parametric (Continuity), as illustrated in Fig.~\ref{fig:sfhpost}.
The posterior distributions were generated using Kernel Density Estimation (KDE) and were not re-normalized to their peaks to preserve differences among the models. 
For the continuity model, we updated the current version 1.2.0 of \textsc{Bagpipes} to incorporate a redshift-adaptive age bin partitioning scheme, enabling analysis of our galaxy sample spanning redshifts $z=1.1-3.4$ (corresponding to a cosmic age range of $5.34-1.81$ Gyr).
Specifically, we employ eight age bins: the first bin encompasses the initial 10 Myr of the look-back time, while the last bin covers the period 100–250 Myr after the Big Bang, corresponding approximately to the epoch from the onset of the first generation of star formation to the emergence of the earliest galaxies. 
The intermediate bins are spaced uniformly in logarithmic time according to the redshift of each galaxy, following \citet{Leja:2019hu}.
Overall, most models yield similar results, with their median values varying less than 0.1 dex, consistent with previous findings \citet{Carnall:2019di, Leja:2019hu}. 
Notably, the Constant SFH generally produces lower mass estimates, while the Continuity SFH results in a slightly higher mass.

We then extended the analysis to the full galaxy sample, 
adopting 3 SFH models—Delayed Tau and Exponential Decline (parametric) as well as Continuity (non-parametric)—to compare the derived $M_*$ with the baseline Double Power Law (DPL) model used in the main text. For further comparison, we also employed the additional independent SED–fitting code \textsc{Beagle} \citep{Chevallard:2016kc}.
In the top panels of Fig.~\ref{fig:sfhoffset}, the 2D KDE (shaded regions) illustrates the mass distribution (small dots) around the one-to-one line (gray lines).
In the bottom panels of Fig.~\ref{fig:sfhoffset}, we calculate the running median of the $M_*$ offset, along with the half-sample medians for subsets divided at $\log(M_*/M_\odot)=8$. 
We can see that for the first Delayed Tau model, due to its similarity in shape to the DPL, including a potentially recent falling phase, results in only minor differences in the derived masses. In contrast, for the Exp. Dec. and Continuity models, although the mass estimates of normal galaxies with $\log(M_*/M_\odot) > 8$ show little variation across different SFH models, the differences can become significant for dwarf galaxies. 
The offset, approximately 0.33 dex, becomes most pronounced at $\log(M_*/M_\odot)\lesssim8.1$ from the running median, closely aligning with \citet{delosReyes:2024arXiv}.
When using different SED fitting codes (e.g., \textsc{Beagle}) with the same choice of SFH, no significant offset is observed across the full mass range; however, local systematic deviations of up to $\sim 0.1$ dex can still occur.
These systematic uncertainties likely arise from the stochastic nature of SFHs in dwarf galaxies at their early stage of evolution before stabilization, which are characterized by short bursts and quenches that simple parameterizations cannot effectively capture.
Therefore, a reliable and accurate determination of the stellar masses of dwarf galaxies requires a more precise understanding and detailed modeling of their SFHs.

\begin{figure*}
  \centering
    \includegraphics[width=\textwidth]{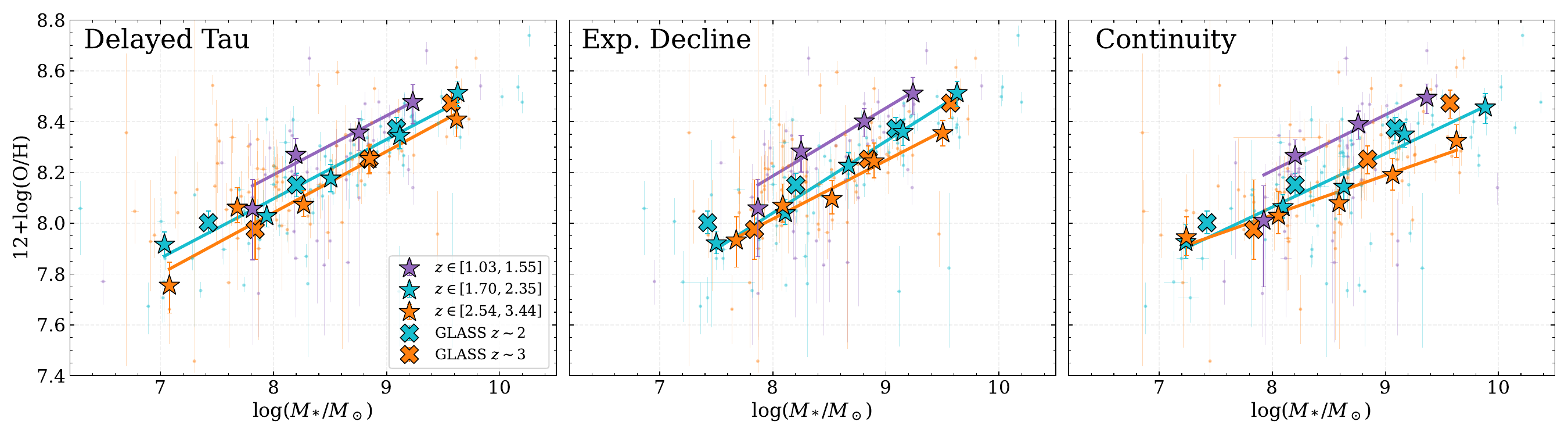}
    \caption{
    Comparison of MZR measurements derived using stellar masses estimated under different SFH models. The legend and plotting format are identical to those in the left panel of Fig.~\ref{fig:mzr}. 
    Each panel shows an independent re-binning and stacking of the galaxies following the procedure described in Sect.~\ref{subsec:stack}. From left to right, the SFH models adopted for the mass estimates are Delayed Tau, Exponential Decline, and Continuity. A summary of the best-fit MZR parameters based on the stacked spectra is provided in Tab.~\ref{tab:mzr}.
    }
    \label{fig:mzr_sfh}
\end{figure*}

\begin{table}
    \centering
    \caption{
        Comparison of the mass--metallicity relation (MZR) from stacked spectra,
        using stellar masses estimated under different SFH models.
        The MZR is defined as
        $12+\log(\mathrm{O/H}) = \beta \times \log(M_*/10^8\,M_\odot) + Z_8$.
    }
    \label{tab:mzr}
    \begin{tabular}{l c c c}
        \hline
        SFH model & $z$ range & slope $\beta$ & intercept $Z_8$ \\
        \hline

           Default:  & 1 & $0.240 \pm 0.014$ & $8.206 \pm 0.013$ \\
           Double Power Law, & 2 & $0.235 \pm 0.017$ & $8.087 \pm 0.018$ \\
           NGDEEP only & 3 & $0.223 \pm 0.029$ & $8.065 \pm 0.030$ \\

        \noalign{\smallskip}\hline\noalign{\smallskip}

          Default:  & 2 & $0.231 \pm 0.020$ & $8.099 \pm 0.019$ \\
          DPL, + GLASS  & 3 & $0.250 \pm 0.031$ & $8.052 \pm 0.030$ \\
          Combination  & 2--3 & $0.238 \pm 0.015$ & $8.078 \pm 0.015$ \\

        \noalign{\smallskip}\hline\noalign{\smallskip}

        \multirow{3}{*}{Delayed Tau}
            & 2 & $0.234 \pm 0.081$ & $8.092 \pm 0.081$ \\
            & 3 & $0.241 \pm 0.114$ & $8.039 \pm 0.106$ \\
            & 2--3 & $0.238 \pm 0.027$ & $8.065 \pm 0.026$ \\

        \noalign{\smallskip}\hline\noalign{\smallskip}

        \multirow{3}{*}{Exp. Decline}
            & 2 & $0.283 \pm 0.047$ & $8.040 \pm 0.045$ \\
            & 3 & $0.230 \pm 0.077$ & $8.015 \pm 0.065$ \\
            & 2--3 & $0.263 \pm 0.026$ & $8.025 \pm 0.024$ \\

        \noalign{\smallskip}\hline\noalign{\smallskip}

        \multirow{3}{*}{Continuity}
            & 2 & $0.210 \pm 0.066$ & $8.062 \pm 0.072$ \\
            & 3 & $0.157 \pm 0.067$ & $8.032 \pm 0.065$ \\
            & 2--3 & $0.188 \pm 0.026$ & $8.046 \pm 0.026$ \\

        \hline
    \end{tabular}
\end{table}

%
Fully incorporating these systematic uncertainties into the estimation of the MZR is a rather complex task. Although we could, in principle, follow approaches such as \citet[][Sect.~5]{Knabel:2025arXiv}, it is impractical to consider all possible SFH models, and it is challenging to combine all the posterior mass distributions from different SFHs and then estimate the stacked median in each mass bin via bootstrap. 
In practice, most current studies still rely solely on the statistical uncertainties of mass estimates under a specific SFH. 
Moreover, our sample benefits from photometry in up to 31 filters, making the statistical uncertainties derived from SED fitting nearly negligible, as illustrated in other figures in the main text (e.g., Figs.~\ref{fig:sfr}-\ref{fig:outflow}). 
Therefore, a more straightforward approach is to directly compare the MZR estimated under different SFHs, providing a qualitative assessment of the systematic uncertainties in mass.
From a practical standpoint, the MZR should be evaluated independently under each SFH assumption rather than adopting the same mass binning defined by the baseline DPL. 
This is because galaxies located near bin edges may shift into adjacent bins when their stellar masses are estimated under different SFHs. Moreover, if rebinning introduces substantial non-uniformities, the original bin edges may also need to be slightly adjusted to ensure that each bin contains a sufficient number of galaxies (e.g., $N > 10$) for robust statistics. Our extensive tests confirm that such minor adjustments to the bin edges do not affect the final MZR estimates.

In Fig.~\ref{fig:mzr_sfh}, we present MZR measurements analogous to those in Fig.~\ref{fig:mzr}, but derived using stellar masses estimated under different SFH models, following the procedure described in Sect.~\ref{subsec:stack}. As in Sect.~\ref{subsec:mzr}, we focus on results from stacked spectra rather than individual galaxies, with the best-fit parameters summarized in Tab.~\ref{tab:mzr}. 
We find that, regardless of the adopted SFH model, dwarf galaxies in our sample exhibit a clear positive correlation between stellar mass and metallicity, indicating that the existence of the MZR is robust to the choice of SFH. 
Moreover, the slope of the MZR remains consistent within $1\sigma$, in the range $\beta\approx 0.22 - 0.24$. While systematic uncertainties in stellar mass due to SFH assumptions can reach $\sim0.3$ dex at $\log(M_*/M_\odot)<8$ (as seen in Fig.~\ref{fig:sfhoffset}), within our mass range of $\log(M_*/M_\odot)=7-10$, this effect impacts only the lowest-mass stacked data point and does not introduce deviations beyond $1\sigma$ in the overall MZR. 
Therefore, the MZR results presented in Sect.~\ref{subsec:mzr}, as well as the subsequent analysis of their physical drivers, remain robust within the quoted $1\sigma$ mass uncertainties induced by SFH.

\subsection{The FMR for dwarf galaxies}\label{subsec:fmr}

\begin{figure}
    \centering
    \includegraphics[width=\columnwidth]{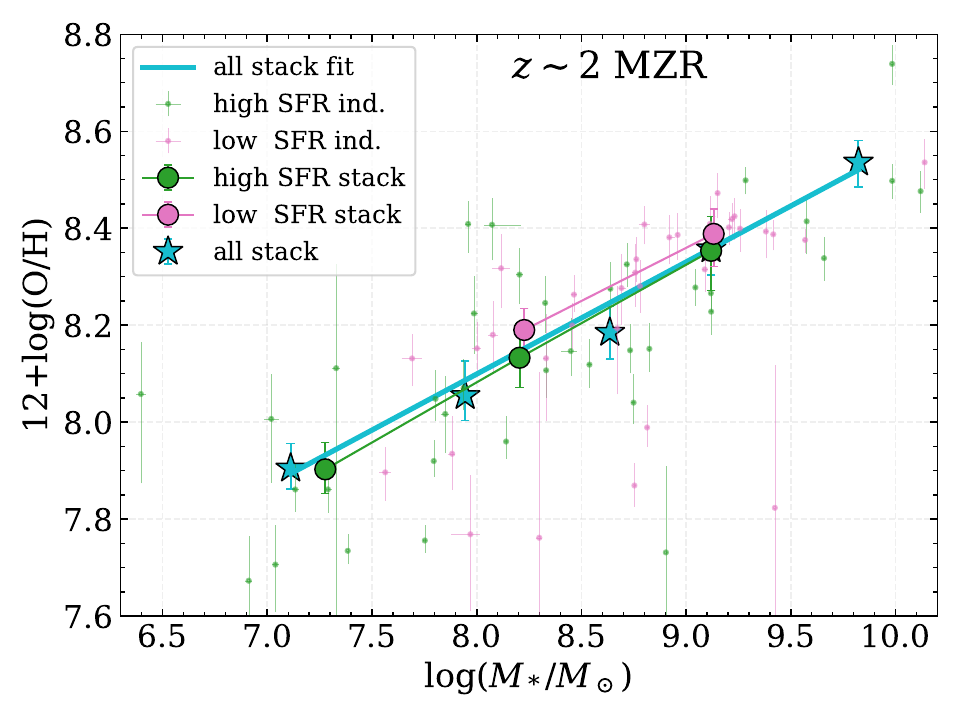}
    \caption{
    MZR measured by splitting the SFR into subsamples at $z\sim2$ as an example, following the method proposed by \citet{Henry:2021ju}.
    The small and large pink dots represent low-SFR individual galaxies and stack results, respectively, while the green dots represent the high-SFR samples.
    The blue stars and the solid line are the results using all low- and high-SFR $z\sim2$ galaxies, identical to those represented by the same color in Fig.~\ref{fig:mzr}.
    }
    \label{fig:sfrbin}
\end{figure}

\begin{figure}  
    \centering
    \includegraphics[width=\columnwidth]{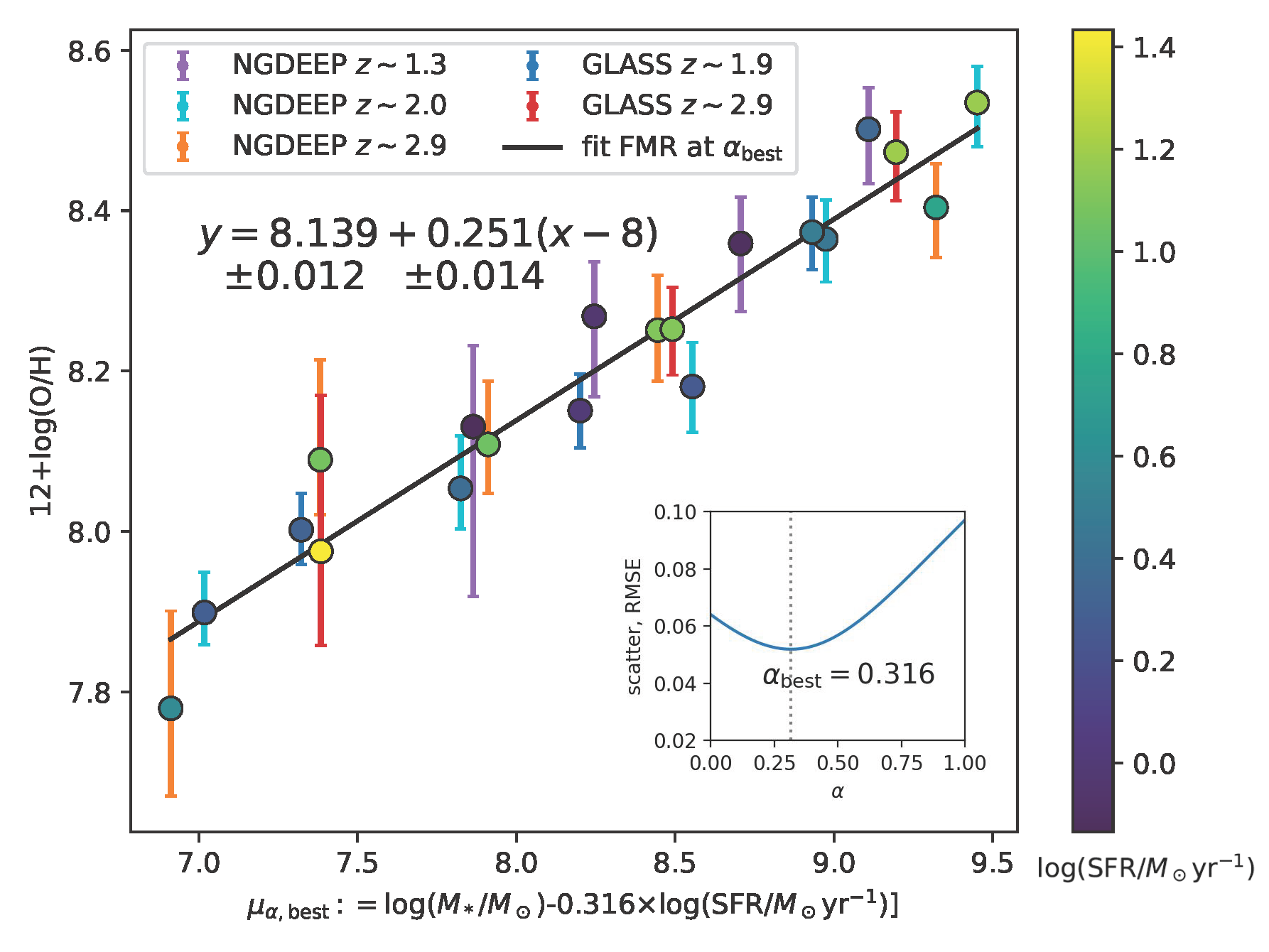}
    \caption{
    Linear projection of the 3-parameter scaling plane FMR, showing 12+log(O/H) as a function of $\mu_\mathrm{best}:= \log(M_*/M_\odot)- a_\mathrm{best}\times\log(\mathrm{SFR}/M_\odot \mathrm{yr}^{-1})$.
    The best coefficient of the SFR $a_\mathrm{best}=0.316$ minimizes the observed scatter (the root mean square error, RMSE) relative to the predicted FMR, as presented in the inset panel. 
    The best-fit FMR with fixed $a$ at minimum RMSE is shown as the black line, and the equation in the upper left.
    Points are color-coded by \Hb-SFR estimated in Sect.~\ref{subsec:metal}, while error bars are color-coded according to their redshift range following Fig.~\ref{fig:mzr}.
    }
    \label{fig:fmr}
\end{figure}


As introduced in Sect.~\ref{sec:intro}, the metallicity \oh is found to be negatively related to the SFR secondarily, creating a three-parameter scaling relation FMR together with $M_*$, for massive galaxies from the local universe 
\citep{Mannucci:2010MNRAS.408.2115M, Andrews:2013dn}.   
In this section, we examine whether current data provide any evidence that such an FMR could extend to dwarf galaxies at $z\sim1-3.5$. 


A direct basis and crucial prerequisite for confirming the local FMR is the direct detection of an SFR dependence in the MZR. 
At high redshifts, however, such attempts are extremely limited \citep[][]{Henry:2021ju}, as typical samples of only tens to hundreds of galaxies render this correlation nearly impossible to detect within the uncertainties. 
This limitation also applies to our dataset, where any putative anti-correlation cannot be discerned across redshift intervals in Fig.~\ref{fig:sfr} and Fig.~\ref{fig:mzr}, i.e., $\Delta Z^8 \propto - \mathrm{SFR}$, as with most of the current research on FMR \citep[e.g.,][]{Curti:2020MNRAS.491..944C}.
To more clearly illustrate the potential SFR dependence of the MZR, we follow the procedure proposed in Sect.~4.4 of \citet{Henry:2021ju}, dividing the 70 galaxies at $z\sim2$ into high- and low-SFR bins and measuring the stacked metallicity. 
The division is based on a linear fit to the SFR–$M_*$ relation at $\log(M_*/M_\odot)>7.5$, with galaxies above (below) the fit classified as high- (low-) SFR  (37 and 33 galaxies, respectively).
As shown in Fig.~\ref{fig:sfrbin}, galaxies with lower SFRs appear to exhibit slightly higher metallicities in the stacked results.
However, this tentative weak anticorrelation is not significant, lying within the $1\sigma$ uncertainties of the stacked metallicities and is insufficient to provide definitive evidence for the existence of the FMR.
Similar to the previous section, we also performed a Chow test on the SFR-distinguished stacked $z\sim2$ MZR, providing a quantitative assessment of the potential subtle correlation between the MZR and SFR.      
The result $F=167.08$ with $p=0.054625$ indicates that the null hypothesis fails to be rejected at the 0.05 significance level, albeit marginally (ignoring the uncertainty of metallicity).
We emphasize that such direct tests of the SFR dependence in the MZR are particularly informative, providing insight into the presence or absence of the FMR. Incorporating similar analyses in future studies could significantly improve the reliability and interpretability of subsequent FMR fitting results, especially at high redshifts.

Despite the weak correlation between the MZR and SFR, we proceed to fit the FMR—by incorporating SFR into the MZR—to examine whether doing so can meaningfully reduce the metallicity scatter.
The FMR is typically expressed as a linear relation between gas-phase metallicity \oh, and a combination of $M_*$ and SFR parameterized as follows:
\begin{equation}
    \mu_a:= \log\left(\frac{M_*}{M_\odot}\right)- a\times\log\left(\frac{\mathrm{SFR}}{M_\odot/\mathrm{yr}}\right).
    \label{eq:mu}
\end{equation}
The free parameter $a$ is introduced to minimize the scatter of \oh for local SDSS galaxies \citep{Mannucci:2010MNRAS.408.2115M}.
It should be noted that the parameters $a,\mu_a$ in this Sect.~\ref{subsec:fmr} are unrelated to the parameters $\alpha,\mu_\mathrm{gas}$ in the previous Sect.~\ref{subsec:mzr}. We use this notation to facilitate direct comparison with results in the literature (e.g., \citet{Mannucci:2010MNRAS.408.2115M} and \citet{Sanders:2021ga}).
Here we take the mainstream approach: scan the parameter space of $a$, calculate the root mean square error (RMSE) relative to the fitted FMR as the scatter at each $a$, and adopt one that minimizes the scatter as the best fit.
This process is shown as the inset panel at the bottom right of Fig.~\ref{fig:fmr}, yielding the relation:
\begin{equation}
\begin{aligned}
    \oh &= \beta\times(\mu_{0.316}-8)+Z^8\\
    \text{with,  } \beta_a &= 0.245\pm0.014, Z^8_a=8.234\pm0.012.
    \label{eq:fmr}
\end{aligned}
\end{equation}
Note that this separate $a_\mathrm{best}$ fit yields some differences from a direct joint $(\beta, Z^8, a)$ fit, because the weights are set the same in RMSE for each point.
We have tested that the differences disappear when setting the weights according to their \oh uncertainties.
We only consider the uncertainty of \oh derived from the stacked spectra, while directly adopting the median $M_*$ and SFR values to construct $\mu_a$. 
The uncertainties in $M_*$ have already been discussed in Sect.~\ref{subsec:sfh}, and the uncertainties in SFR will be addressed in detail in Sect.~\ref{subsec:sfrerr}.

The best-fit parameter $a$ reduces the scatter to $\mathrm{RMSE}=0.052$ for $a=0.316$, compared to $\mathrm{RMSE}=0.064$ for $a=0$ (corresponding to the unevolved MZR) and $\mathrm{RMSE}=0.097$ for $a=1, \mu_1=-\log\mathrm{sSFR}$ \citep[as established by][as a precursor]{Ellison:2008dy}. 
For reference, the reduction of scatter in the local FMR is $\mathrm{RMSE} = 0.057 \rightarrow 0.021$ at $a=0.32$ \citep{Mannucci:2010MNRAS.408.2115M}. 
While our reduction in scatter is formally measurable, it is far from significant: typical uncertainties in the stacked metallicities exceed 0.05 dex, rendering the apparent improvement marginal at best. 
Similar weak or negligible correlations have been reported in high-redshift studies \citep[e.g.,][$0.17\rightarrow0.16$ at $a=0.17$]{Henry:2021ju} and in other recent work \citep[e.g.,][reduction of 0.004 dex]{Stephenson:2024MNRAS}.
Furthermore, as shown in Fig.~\ref{fig:sfrbin}, the modest excess of low-SFR over high-SFR stacked metallicities ($\Delta\log(\mathrm{O/H})\approx0.045$) is consistent with the prediction from our fit ($\beta_a \cdot a \cdot \Delta(\mathrm{logSFR}_\mathrm{median})\approx0.245\times0.316\times0.6=0.046$). However, given the size of the uncertainties, this agreement does not constitute statistically significant evidence for the presence of the FMR. In short, while the trend is suggestive, our current dataset does not provide robust support for the FMR at these redshifts.

Previous studies report a range of correlation strengths (if assumed to exist), from the weaker branch $a\sim0.17-0.32$     \citep{Mannucci:2010MNRAS.408.2115M,Yates:2012kx,Guo:2016ApJ,Hirschauer:2018AJ,Henry:2021ju,Pharo:2023ApJ} to stronger relations $a\sim0.6$     \citep{Andrews:2013dn,Sanders:2021ga,Li:2023ApJL,Langan:2023MNRAS,Stephenson:2024MNRAS}, highlighting the variability introduced by different methods, metallicity calibrations, and galaxy samples. Nonlinear three-parameter formulations have also been proposed \citep{Curti:2020MNRAS.491..944C}.
The dependence of the MZR on the SFR at high redshift and dwarf galaxies has been under debate \citep{Maiolino:2019vq,Kewley:2019kf}, and thus much of the other recent works stop at comparing and discussing their deviation relative to the local FMR \citep{Heintz:2023NatAs,Curti:2023MNRAS,Curti:2023arXiv230408516C,Langeroodi:2023arXiv, Llerena:2023A&A, Bulichi:2023A&A, Sarkar2024arXiv240807974S}.
Taken together, these results emphasize that current observations at cosmic noon do not provide robust support for a universal FMR, and any apparent SFR dependence should be interpreted with caution.
\citet{Henry:2021ju} argues that some of the discrepancy may come from the correlation between SFR and strong-line calibration.
\citet{Langan:2023MNRAS} interprets this SFR dependence as a strong inflow promoting star formation activity and diluting metallicity.
If the stronger correlation $a\simeq0.6$ for the local FMR is adopted, our best-fit value of $a\simeq0.3$ at $z\sim2$ could in principle be interpreted as a weaker high-redshift counterpart, although marginally. 
While this weaker trend could be qualitatively attributed to the shallow potential wells of dwarf galaxies, where feedback-driven outflows more effectively suppress the impact of SFR on metallicity, such an interpretation should be regarded as speculative.

\rp{
A similar picture emerges from numerical simulations.
Cosmological simulations generally predict that, at fixed stellar mass, galaxies with enhanced recent star formation tend to have lower gas-phase metallicities, producing a secondary SFR dependence of the MZR \citep{DeRossi:2017ge,Torrey:2018he,Garcia:2024arXiv}.
Physically, this behaviour is often interpreted as the response of the ISM to recent gas accretion, dilution, star formation, and feedback-regulated metal loss.
However, simulations also suggest that the strength and redshift evolution of this correlation are model-dependent.
For example, \citet{Torrey:2018he} showed that an FMR requires SFR and metallicity residuals to evolve on comparable timescales, while bursty star-formation histories can weaken the residual correlation.
More recently, \citet{Garcia:2024arXiv} found that Illustris, IllustrisTNG, and EAGLE predict a secondary SFR dependence up to high redshift, but with an evolving correlation strength rather than a strictly redshift-invariant FMR.
Therefore, the absence of a statistically robust FMR signal in our low-mass NIRISS sample does not necessarily contradict theoretical expectations; instead, it may reflect the combination of bursty star formation, calibration uncertainties, and the limited precision of current metallicity and SFR measurements in high-redshift dwarf galaxies.
}


More importantly, our conclusion is consistent with recent studies questioning whether the local FMR can be directly extended to high redshift or to the low-mass regime.
At a similar redshift $z\sim2.3$, \citet{KorhonenCuestas:2025ApJ} conducted a systematic study of the FMR for KBSS galaxies with $\log(M_*/M_\odot)\in[9,11]$, and found no robust evidence for a direct extension of the local FMR to high redshift, regardless of the adopted methodology.
\rp{
At lower masses, \citet{Laseter:2025arXiv} showed that even the local FMR is not well established below $\log(M_*/M_\odot)\lesssim9$, finding no significant FMR signal in a large direct-$T_{\rm e}$ sample of low-mass galaxies.
Our study therefore provides a complementary high-redshift test in the dwarf-galaxy regime, i.e., $z\sim2$ and $\log(M_*/M_\odot)\in[7,9]$.
}
Overall, whether a robust FMR exists for high-redshift dwarf galaxies remains an open question.
Our measurements add useful constraints in this direction, but should be interpreted primarily as a step toward establishing a firmer low-mass MZR benchmark at cosmic noon, rather than as definitive evidence for an FMR.

\subsection{Systematic Uncertainties of SFR}
\label{subsec:sfrerr}

\begin{figure*}
    \centering
    \includegraphics[width=\textwidth]{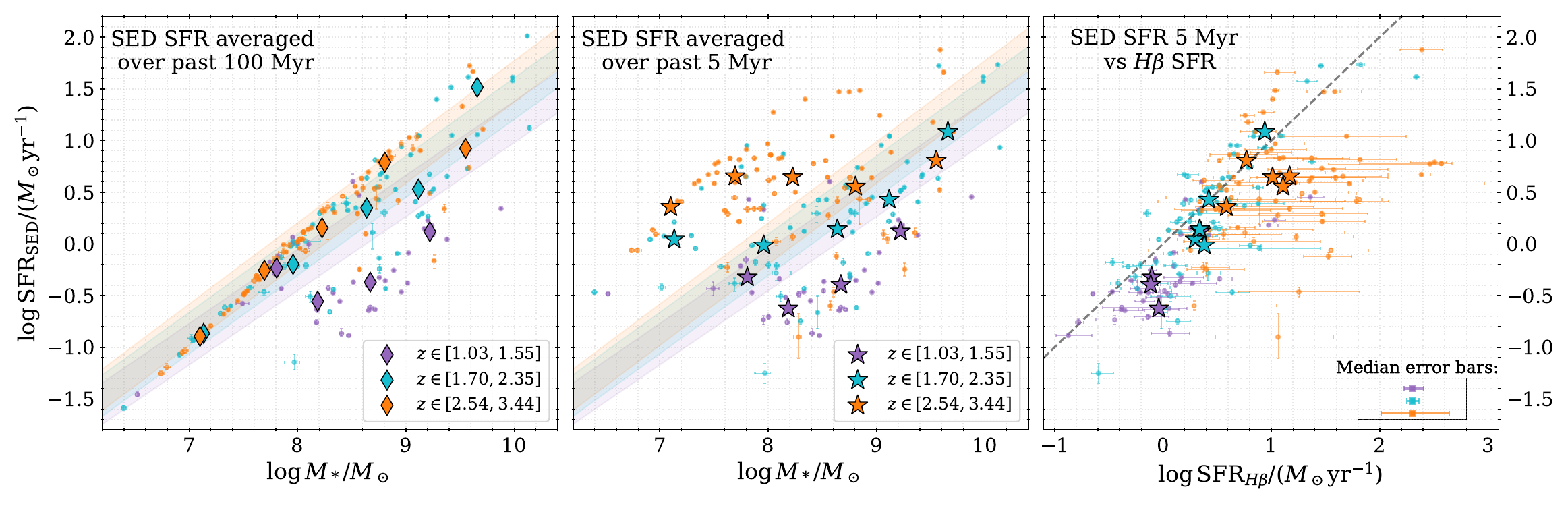}
    \caption{
   \textit{Left:} the SFR--$M_*$ relation of our sample, similar to Fig.~\ref{fig:sfr}, but with H$\beta$-based SFR replaced by SED-based SFR derived simultaneously with $M_*$ in Sect.~\ref{subsec:mass}. In common SED-fitting codes (e.g., \textsc{Bagpipes}), the SFR corresponds to the average star-formation rate over the latest $100$ Myr. 
   \textit{Middle:} the SFR--$M_*$ relation similar to the left panel, but with SED-based SFR measured over the most recent 5 Myr. This instantaneous tracer is more comparable to the H$\beta$-based SFR, and thus is adopted as the default choice in our FMR analysis in Sect.~\ref{subsec:sfrerr}. 
   \textit{Right:} comparison between H$\beta$-based SFR (x-axis) and SED-based SFR on a 5 Myr timescale (y-axis) for individual galaxies and their medians in the mass bins of Sect.~\ref{subsec:stack}. We also show the corresponding median error bars for individual galaxies in each redshift range at the bottom left box, together with the one-to-one relation (dashed gray line). 
   Note that the median uncertainties of all SED-derived quantities here, including $M_*$ and both SFR estimators, are in $0.01-0.015$ dex, well below the marker size and therefore not visible, in contrast to the larger uncertainties of the H$\beta$-based SFR.
}
    \label{fig:sfr_sed5}
\end{figure*}

In the preceding section on the FMR, we found that even when neglecting the uncertainties in $M_*$ and SFR in Eq.~\ref{eq:mu}, no robust evidence for the existence of an FMR could be obtained. For completeness, however, it is still necessary to examine the impact of SFR uncertainties, which is the focus of this section.

The SFR used for stacked spectra in Eq.~\ref{eq:mu} is taken as the median of the individual galaxies, which is based on \Hb (rather than SED), and simultaneously estimated with $A_\nu$ and \oh\ in Sect.~\ref{subsec:metal}. 
In contrast, the SFR cannot be directly estimated from the stacked H$\beta$, since prior to stacking each spectrum has been normalized by its \OIII flux. This procedure removes the absolute flux information, retaining only relative flux ratios for metallicity estimates. 
\citet{Sanders:2021ga} attempted to recover absolute fluxes by multiplying the relative stacked H$\beta$ with the median \OIII flux of individuals used for normalization. An alternative approach, adopted by \citet{Matthee:2023ApJ}, is to construct luminosity stacks, where each spectrum is rescaled by its luminosity distance (a function of redshift), which may better reflect the intrinsic properties. 
We applied both methods to extract the absolute flux $f^o_{\Hb}$, corrected for dust attenuation following Sect.~\ref{subsec:metal} to obtain $f_{\Hb}$ and the corresponding stacked H$\beta$-based SFR.
In both cases, the stacked-\Hb SFRs are systematically lower than the median individual-\Hb SFRs,
\rp{
with a median offset of $\Delta\log{\rm SFR}\simeq -0.32$ dex and a range from $-0.62$ to $+0.02$ dex, as listed in Tab.~\ref{tab:stack}.
}
This systematic underestimation is not a coincidence, but is consistent with the morphological broadening effects discussed by \citet{He:2024ApJL}. 
Indeed, stacking 1D spectra without rescaling can only provide unbiased relative line ratios, whereas absolute line fluxes (e.g., $f_{\Hb}$) require stacking 2D spectra. A detailed comparison of different stacking strategies is beyond the scope of this work, and we therefore adopt the median of individual SFRs as our fiducial choice for FMR analysis in Sect.~\ref{subsec:fmr}.

In addition to the H$\beta$-based SFR, Sect.~\ref{subsec:mass} provides simultaneous estimates of SFR from SED fitting together with $M_*$. The choice of tracer itself may introduce systematic differences, which we investigate in the latter part of this section. 
SED-based SFRs typically represent averages over relatively long timescales \citep[e.g., $100$ Myr,][]{Carnall:2018gb}, since photometric fluxes are the integrated light of stellar populations.
This is analogous to the UV+IR tracers commonly adopted in local SFMS studies, which are sensitive to stars formed within the past 10–100 Myr. As a result, when we replot the SFR–$M_*$ relation using SED-derived SFRs (left panel of Fig.~\ref{fig:sfr_sed5}), the sequence lies much closer. 

For the FMR, however, what matters is the instantaneous SFR, usually traced by \Hb emission on $3-10$ Myr timescales \citep{Kennicutt:2012ey}. The physical motivation is that elevated SFRs correspond to enhanced gas inflows, which dilute the ISM metallicity and lower $Z$, thus emphasizing short-term gas accretion and star formation. 
To approximate this, we modified the \textsc{Bagpipes} setup in Sect.~\ref{subsec:mass} to compute SED-based SFRs on a 5 Myr timescale. The resulting SFR–$M_*$ relation shown in the middle panel of Fig.~\ref{fig:sfr_sed5}, illustrates systematically higher SFRs than the 100 Myr estimates, indicating that the majority of our galaxies are undergoing vigorous star formation at the observed redshifts. 
As expected, the 5 Myr SED SFRs closely follow the H$\beta$-based relation as Fig.~\ref{fig:sfr}, and their direct comparison (right panel of Fig.~\ref{fig:sfr_sed5}) further demonstrates this similarity. 
The main deviations occur at $z\sim3$, primarily due to lower SNR and the lack of H$\alpha$, which hampers accurate estimates of $A_V^\mathrm{gas}$ and thus \Hb SFR. 
While in principle one could adopt even shorter averaging timescales to better approximate the instantaneous H$\beta$ SFR under the "rising" SFH, a balance must be struck between capturing bursty behavior and preserving the stability of SFH-averaged quantities.



Using SFRs derived from SED fitting with a short timescale of 5 Myr above, we follow the approach of Sect.~\ref{subsec:sfh} and repeat the exploration of the FMR presented in Sect.~\ref{subsec:fmr}. 
Unfortunately, we still find no clear evidence supporting the existence of the FMR. 
This holds both for the analysis shown in Fig.~\ref{fig:sfrbin}, where galaxies are classified into high- and low-SFR bins and then stacked to compare deviations in metallicity, and for the approach in Fig.~\ref{fig:fmr}, where the inclusion of an SFR parameter is intended to minimize the metallicity scatter. 
In both cases, the statistical uncertainties in metallicity continue to obscure any potential evidence. 
It should be noted that the SED-based SFRs are also subject to systematic uncertainties similar to those affecting the $M_*$ measurements discussed in Sect.~\ref{subsec:sfh}. 
These uncertainties arise from the choice of SFH model, in addition to the tracer used (H$\beta$ or SED) and the adopted timescale (5 or 100 Myr), further limiting our ability to identify clear evidence for the FMR.

Therefore, robust confirmation of the FMR requires more accurate modeling and calibration. In addition to metallicity, precise measurements of both stellar mass and SFR, along with a careful assessment of all potential systematic uncertainties, are crucial. 
Given the current observational precision of NIRISS, confirming the FMR in high-redshift dwarf galaxies remains extremely challenging. Nevertheless, our exploration using the largest sample to date provides tentative indications and highlights directions for future investigations.


\section{Summary and Conclusions }\label{sec:conclude}

We have presented measurements of the MZR and investigated the possible existence of an FMR for dwarf galaxies with $M_*\in(10^{6.3},10^{10.2})\,M_{\odot}$ at cosmic noon ($z\in[1.1,3.4]$) using JWST/NIRISS slitless spectroscopy.
We perform an initial analysis of the publicly available first-epoch NGDEEP grism data targeting the HUDF. 
The spectroscopically identified sources are matched to the public DJA photometric catalogue.
An unprecedented sample of 183 field galaxies is selected in three redshift ranges $z\in [1.10, 1.54], [1.76,2.32], [2.61,3.43]$, with measured $M_*/M_\odot \in (10^{6.3},10^{10.2})$ from SED fitting.
Their spectra are grouped and stacked according to their redshift and mass in order to obtain higher SNR emission line flux. 
We employ our forward modeling Bayesian method to infer the metallicity for both the individual galaxies and the stacked line fluxes, assuming the strong line calibration of \citet{Bian:2018km}.

The MZR derived from the NGDEEP field aligns very well with our previous results from the GLASS-JWST field \citep{He:2024ApJL}, but with a threefold increase in sample size and an extension to lower masses.
After combining the GLASS sample, the stacked MZR can be characterized by: $12 + \log(\mathrm{O/H}) = \beta \times\log(M_*/10^8M_\odot)+ Z^8$ with 
$\beta= 0.240\pm0.014,0.231\pm0.019,0.250\pm0.031\text{, at } z_\mathrm{med}=1.304,2.022,2.947$, respectively, and a slight $Z^8$ evolution conducted by the Chow test, as presented in Fig.~\ref{fig:mzr} and Eq.~\ref{eq:mzr_def}.
These results once again strengthen our previous conclusion that the MZR (e.g. \citealt{Sanders:2021ga}) can be extrapolated quite reliably down to the low mass end of $M_*\sim10^7M_\odot$ at cosmic noon, though with larger scatter.
Nevertheless, we likewise find a trend for dwarf galaxies to exhibit a slightly shallower MZR, 
\rp{
as also suggested by previous low-mass studies at intermediate and high redshift (e.g., \citealt{Guo:2016ApJ,Henry:2021ju,Curti:2023arXiv230408516C,Revalski:2024arXiv}).
}
This may indicate that the relative importance of different feedback prescriptions, such as energy-driven and momentum-driven winds, changes toward the low-mass regime.
As an example, we calculate the metal loading factor of outflow $\zeta_\text{out}$ following the model performed in \citet{Sanders:2021ga}.
We find that the gas-fraction term $\alpha\cdot\mu_\mathrm{gas}$ gradually gains importance toward lower masses, becoming comparable to or larger than the metal-loading term $\zeta_\mathrm{out}$ at $M_*\lesssim10^8M_\odot$ for $z\gtrsim2$, as shown in Fig.~\ref{fig:outflow}.
\rp{
This suggests that the gas reservoir is a key ingredient in setting the low-mass MZR normalization at cosmic noon.
However, this should not be interpreted as evidence that pristine-gas dilution alone dominates the metallicity evolution of dwarf galaxies.
Recent effective-yield analyses of low-mass \OIII$\lambda4363$ emitters suggest that, although such systems are gas rich, their metallicity diversity is more closely connected to bursty star formation and metal-enriched outflows than to gas fraction alone \citep{Laseter:2025arXiv}.
}

\rp{
We have also quantified the impact of two main sources of systematic uncertainty on the MZR: the metallicities \oh inferred using different strong-line calibrations, and the stellar masses $M_*$ inferred under different SFH assumptions.
For the metallicity-calibration systematics shown in Fig.~\ref{fig:mzr_calib} and Tab.~\ref{tab:mzr_calib}, different calibrations can shift the absolute metallicity scale by up to $\Delta Z\sim0.5$ dex at fixed stellar mass, but the fitted MZR slopes vary only modestly in most cases, with $\beta\simeq0.23-0.27$.
For the stellar-mass systematics shown in Fig.~\ref{fig:mzr_sfh} and Tab.~\ref{tab:mzr}, although different SFH assumptions can shift individual masses by up to $\sim0.3$ dex at only the lowest masses, the resulting MZR slopes remain consistent within $1\sigma$, with $\beta\simeq0.22-0.24$.
Thus, while these systematics do not alter the existence of a positive MZR in our low-mass cosmic-noon sample, they remain important for quantitative comparisons of the MZR normalization and slope with other studies.
}

In addition, we find no compelling evidence for extending the locally proposed FMR to dwarf galaxies at cosmic noon. As shown in Fig.~\ref{fig:sfrbin}, the offset between the high- and low-SFR stacks is marginal and well within the uncertainties, offering at best tentative hints for an SFR dependence in the MZR. Likewise, the best-fit coefficient $a_\text{best}=0.316$ (Fig.~\ref{fig:fmr}, Eq.~\ref{eq:fmr}) yields only a negligible reduction in scatter relative to the dominant metallicity uncertainties.
This conclusion should also be interpreted in light of the metallicity-calibration systematics discussed in Sect.~\ref{subsec:calib}.
Because different strong-line calibrations can shift the absolute metallicity scale by up to $\sim0.5$ dex and change the detailed MZR slope, the current data are not sufficient to establish a residual SFR-dependent metallicity variation at the $\lesssim0.1$ dex level as a robust FMR signal.
These findings suggest that the shallow potential wells of dwarf galaxies may suppress the influence of star formation on chemical enrichment, or that such systems have not yet reached the equilibrium state at cosmic noon, where an FMR would naturally emerge. 
\rp{
We also explored the impact of systematic uncertainties in SFR on the FMR analysis as in Fig.~\ref{fig:sfr_sed5}.
Together with the systematic uncertainties in \oh and $M_*$, uncertainties in SFR further blur any underlying secondary correlation, effectively hiding metallicity variations at the $\lesssim0.1$ dex level.
Thus, while our data do not rule out a weak FMR in high-redshift dwarf galaxies, robust confirmation of such a relation requires more precise and consistently calibrated measurements of $\oh$, $M_*$, and SFR.
}

\rp{
Our study complements recent work in both the massive and dwarf-galaxy regimes.
\citet{KorhonenCuestas:2025ApJ} found no evidence for a statistically significant FMR at $z\sim2.3$ among more massive galaxies, while \citet{Laseter:2025arXiv} argued that the canonical FMR may break down below $\log(M_*/M_\odot)\lesssim9$ in local and high-redshift dwarf galaxies.
Taken together, these results point toward a consistent picture in which the SFR dependence of the MZR at high redshift is weak, sample-dependent, and potentially blurred by non-equilibrium gas cycling in low-mass galaxies.
}
Looking ahead, our early use of the first NGDEEP epoch data provides a valuable starting point and observational benchmarks. 
With forthcoming surveys delivering larger, deeper, and more precise datasets, it will become possible to revisit the question of whether an FMR can be meaningfully defined in the early universe.


\section*{Acknowledgements}

X. H. thanks Xiaolei Meng, and Mingyu Li for useful discussions.
X. H. is supported by the China Scholarship Council. 
X. W. is supported by the China Manned Space Program with grant no. CMS-CSST-2025-A06, the National Natural Science Foundation of China (grant 12373009), the CAS Project for Young Scientists in Basic Research Grant No. YSBR-062, and the Fundamental Research Funds for the Central Universities.
X. W. also acknowledges the support by the Xiaomi Young Talents Program, the work carried out, in part, at the Swinburne University of Technology, sponsored by the ACAMAR visiting fellowship, and the hospitality of the International Centre of Supernovae (ICESUN), Yunnan Key Laboratory at Yunnan Observatories Chinese Academy of Sciences.
This work is based on observations made with the NASA/ESA/CSA James Webb Space Telescope.  Support by NASA through grant JWST-ERS-1324 is gratefully acknowledged.
The data were obtained from the Mikulski Archive for Space Telescopes at the Space Telescope Science Institute, which is operated by the Association of Universities for Research in Astronomy, Inc., under NASA contract NAS 5-03127 for JWST. 
(Some of) the data products presented herein were retrieved from the Dawn JWST Archive (DJA). DJA is an initiative of the Cosmic Dawn Center (DAWN), which is funded by the Danish National Research Foundation under grant DNRF140.
We also acknowledge support from the INAF Large Grant 2022 “Extragalactic Surveys with JWST” (PI Pentericci). BV is supported by the European Union – NextGenerationEU RFF M4C2 1.1 PRIN 2022 project 2022ZSL4BL INSIGHT.




We thank the use of JWST (NIRISS, NIRCam) and HST (ACS, WFC3) for the observations analyzed in this work.  
Our analysis made use of the following software: \textsc{Grizli} \citep{Brammer:2021df}, \textsc{Bagpipes} \citep{Carnall:2018gb}, \textsc{LMfit} \citep{Newville:2021cv}, \textsc{Emcee} \citep{ForemanMackey:2013io}, SciPy \citep{2020SciPy-NMeth}, NumPy \citep{harris2020array}, Matplotlib \citep{Hunter:2007}, and Astropy \citep{astropy:2022}.


\section*{Data availability}

The JWST NIRISS WFSS observations analyzed in this work originate from the first-epoch NGDEEP survey \citep[Proposal ID \#2079;][]{Bagley:2023arXiv} and were retrieved from the public MAST archive. The specific calibrated data products used in this study correspond to the release available under DOI \href{https://doi.org/10.17909/78mb-cv98}{10.17909/78mb-cv98}.
The reduced spectra and emission-line measurements produced via \textsc{Grizli} are available from the authors on reasonable request. 
The NIRCam photometry catalog for the NGDEEP field is publicly accessible from The DAWN JWST Archive (\href{https://dawn-cph.github.io/dja/imaging/v7/}{DJA}).



\bibliographystyle{mnras}
\bibliography{reference,ngdeep}{}



\appendix 

\renewcommand{\thetable}{A\arabic{table}}
\renewcommand{\thefigure}{A\arabic{figure}}
\setcounter{table}{0}  
\setcounter{figure}{0}
\section{Emission line sensitivity}
\label{sec:appendix}

\begin{figure*}
    \centering
    \includegraphics[width=\textwidth]{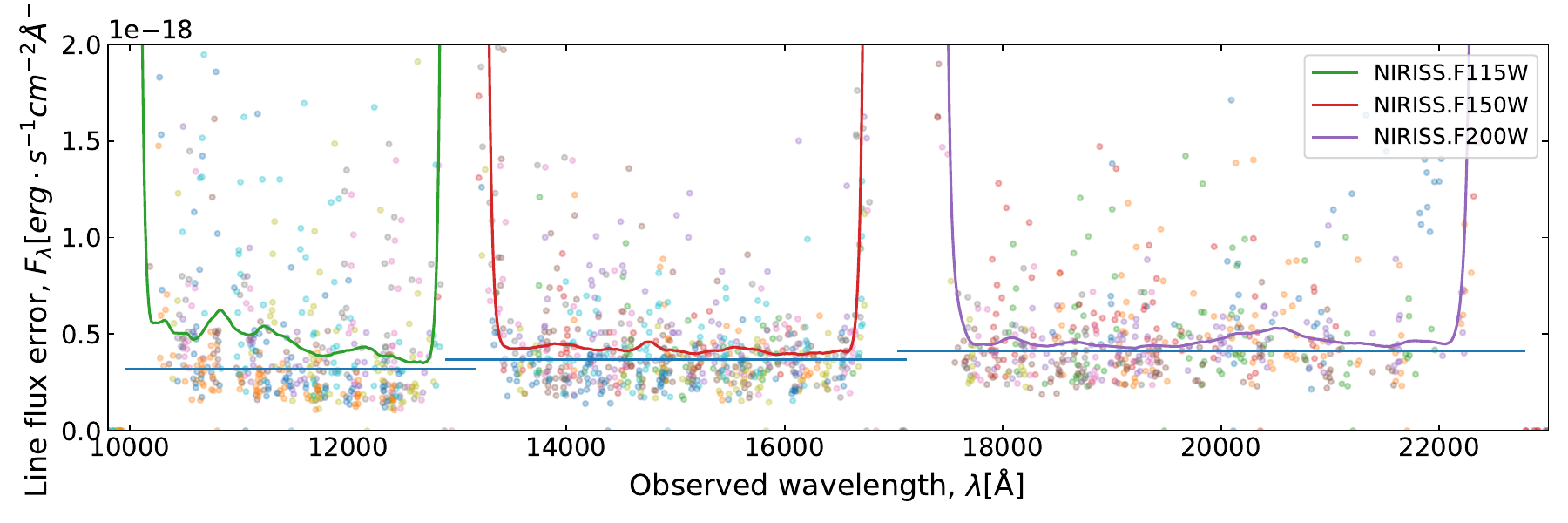}
    \caption{
    The emission line sensitivity of our sample. The color-coded scatters represent the measured error of each emission line ( SIII,  
    \SII, \Ha, \OI-6302, \OIII, \Hb, \OIII-4363, \Hg, \Hd, \NeIII-3867, \OII,  MgII ). The 3 blue horizontal lines mark the median error in 3 bands with $F_\lambda \in (0, 10^{-18})$, which is considered to be the $1\sigma$ line sensitivity for all the lines.
    The line limit of 0.3 is as expected by \citet{Bagley:2023arXiv}.
    Therefore, the $3\sigma$ detections of \Hb are used to label the SFR limit.
    As presented in \citet{Momcheva:2016fr}, the line uncertainty is inversely proportional to the second power of each grism throughput: $\sigma \propto \mathrm{(throughput)}^{-2}$, after being normalized to their median.
    }
    \label{fig:flux_limit}
\end{figure*}

Here we discuss how to derive the emission line sensitivity of our sample, the 3 horizontal lines in Fig.~\ref{fig:sfr}. 
We use the emission line errors measured by \grzl for almost all lines to characterize the flux limit of the observation.
We plot the line error versus observed wavelength in Fig.~\ref{fig:flux_limit}.
We also show the negative quadratic power of the grism throughput, which is expected to be proportional to the flux limit  \citep[Fig.19,][]{Momcheva:2016fr}.
We counted the median for all line errors that were less than $10^{-18}\mathrm{erg}\cdot \mathrm{s}^{-1}\mathrm{cm}^{-2}\AA^{-1}$ and non-zero, divided by filters, and regarded them as the $1\sigma$ detection limits.
The median $1\sigma$ errors are $0.32, 0.37, 0.41 \times 10^{-18}$ for NIRISS F115W, F150W, F200W, respectively, consistent with $5\sigma$ expectations of \citet{Bagley:2023arXiv}.
The SFR limits are calculated from these $3\sigma$ detection limits following Eq.~\ref{eq:sfr}, with the result of $0.13, 0.42, 1.28 {M_\odot\mathrm{\,yr^{-1}}}$ marked in Fig.~\ref{fig:sfr}.

\label{lastpage}
\end{document}